\documentclass[useAMS,usenatbib]{mn2e}

\usepackage{amssymb}
\usepackage{graphicx}

\title[Outburst evolution of Swift J1753.5-0127]{A complex state transition from the black hole candidate Swift J1753.5-0127}
\author[P. Soleri et al.]{P. Soleri$^{1}$\thanks{E-mail:
soleri@astro.rug.nl}, T. Mu{\~n}oz-Darias$^{2,3}$, S. Motta$^{4,3}$, T. Belloni$^{3}$, P. Casella$^{5}$, M. M{\'e}ndez$^{1}$,\newauthor
D. Altamirano$^{6}$, M. Linares$^{7}$, R. Wijnands$^{6}$, R. Fender$^{2}$ and M. van der Klis$^{6}$\\
$^{1}$Kapteyn Astronomical Institute, University of Groningen, PO Box 800, 9700 AV, Groningen, The Netherlands\\
$^{2}$School of Physics and Astronomy, University of Southampton, Hampshire, SO17 1BJ, UK\\
$^{3}$INAF-Osservatorio Astronomico di Brera, Via E. Bianchi 46, I-23807 Merate  (LC), Italy\\
$^{4}$European Space Astronomy Centre (ESAC)/ESA, PO Box 78, E-28691 Villanueva de la Ca{\~n}ada, Madrid, Spain\\
$^{5}$INAF-Osservatorio Astronomico di Roma, via Frascati 33, Monteporzio Catone, 00040, Italy\\
$^{6}$Astronomical Institute Anton Pannekoek, University of Amsterdam, Science Park 904, 1098 XH, Amsterdam, The Netherlands\\
$^{7}$Instituto de Astrof{\'i}sica de Canarias (IAC), V{\'i}a L{\'a}ctea s/n, La Laguna, E-38205, S/C de Tenerife, Spain
}
\begin{document}

\date{Accepted 2012 November 14. Received 2012 November 13; in original form 2011 October 14}

\pagerange{\pageref{firstpage}--\pageref{lastpage}} \pubyear{yyyy}

\maketitle

\label{firstpage}

\begin{abstract}
%
We present our monitoring campaign of the outburst of the black-hole candidate Swift~J1753.5-0127,
observed with the Rossi X-ray Timing Explorer and the Swift satellites. After $\sim 4.5$
years since its discovery, the source had a transition to the hard intermediate state.  
We performed spectral and timing studies of the transition showing that, unlike the majority of the
transient black holes, the system did not go to the soft states but it returned to the hard state after
a few months. During this transition Swift J1753.5-0127 features properties which are similar to those displayed by the
black hole Cygnus X-1. We compared Swift J1753.5-0127 to one dynamically confirmed black hole
and two neutron stars showing that its power spectra are in agreement with the binary hosting a black hole.
We also suggest that the prolonged period at low flux that followed the initial flare is reminiscent 
of that observed in other X-ray binaries, as well as in cataclysmic variables.
\end{abstract}


\section{Introduction} \label{par:intro}
Black hole X-ray transients (BHTs) are black hole X-ray binaries which usually show relatively short (week to months) outbursts,
separated by long periods of quiescence. During an outburst, both timing and spectral properties vary. These changes are used to define
different states (see Belloni, Motta \& Mu{\~n}oz-Darias 2011 for a review).
In the low/hard state (LHS), observed at the beginning and at the end of the outburst,
the X-ray spectrum is dominated by a component which can be approximated by a power law with a photon index of
$\sim 1.5-1.8$ and a high-energy cut-off at $\lesssim 200$ keV. The power density spectrum (PDS) in the LHS features a high level of aperiodic
variability with a fractional root mean square (rms) amplitude usually above $\sim 30 \%$ and often a quasi-periodic oscillation (QPO).
In the high/soft state (HSS) the aperiodic variability drops to values of few per cent and the energy spectrum becomes softer
and dominated by a thermal disc-blackbody component. The HSS is usually observed in the middle of the outburst.
In between these two states the situation is more complex, with hard-to-soft and soft-to-hard transitions usually taking place on
relatively short time-scales (hours to days) and associated with dramatic changes in the timing and spectral properties.
Two additional states can be identified, the hard-intermediate state (HIMS) and the soft-intermediate state (SIMS).

A widely used tool to study the spectral evolution of BHTs in outburst is the X-ray hardness-intensity diagram (HID),
in which different spectral states correspond to different branches/areas of a q-shaped HID pattern (Homan et al. 2001).
During an outburst, BHTs are usually observed in all the four states defined above. However, ``hard outbursts'' are also observed
(Brocksopp, Bandyopadhyay \& Fender 2004): in this case the BHT spends the whole outburst in the LHS, or in the LHS and in the HIMS
(e.g. H~1743-322;  Capitanio et al. 2009) without transiting to the soft states. It is worth noting that some sources underwent
both ``normal'' outburst and hard outbursts (i.e., with or without a transition to the soft states, respectively;
e.g. XTE~J1550-564, Belloni et al. 2002b).
%

X-ray variability is a good indicator of the spectral state of a source: for instance Uttley \& McHardy (2001)
discovered that black holes in the LHS feature a linear relation between the absolute rms amplitude of the variability and count rate.
Following the evolution of the count rate and the rms, Mu{\~n}oz-Darias et al. (2011) developed a new tool to map the states of a BHT in
outburst: the rms-intensity diagram (RID). In this diagram, the linear relation between the absolute rms and the count rate is represented by
the so-called hard line. Mu{\~n}oz-Darias et al. (2011) used the RID to study three outbursts of the BHT GX 339-4, showing that it is
possible to associate different regions of the diagram to different states. 
\begin{figure*}
\begin{tabular}{c}
\resizebox{16.8cm}{!}{\includegraphics{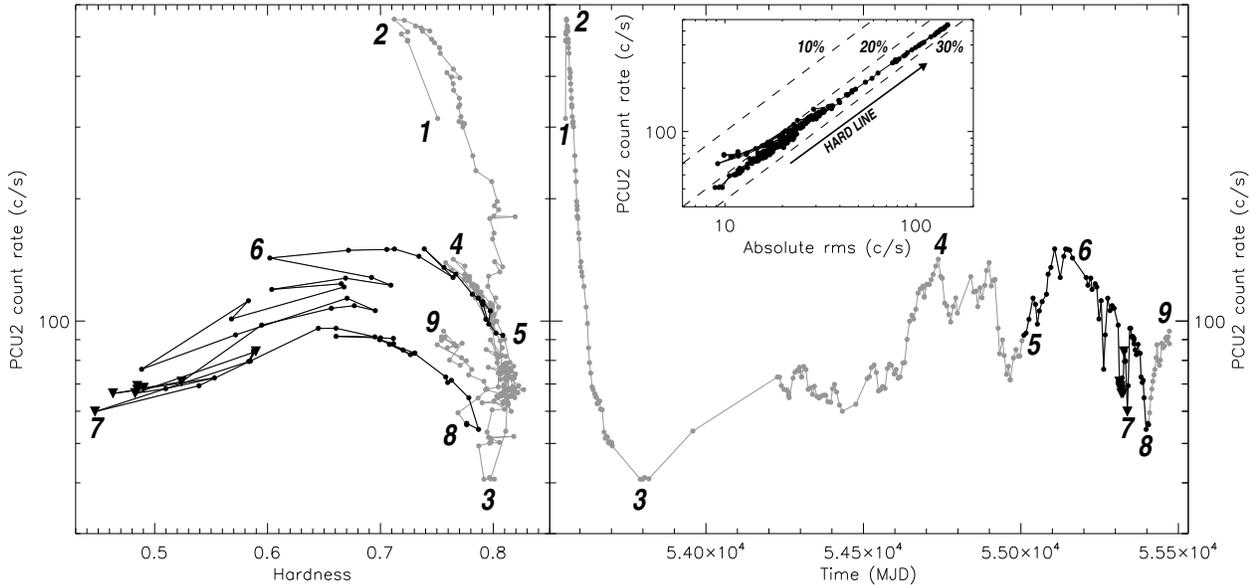}}
\end{tabular}
\caption{X-ray hardness-intensity diagram and light curve of Swift J1753.5-0127 (left and right panel, respectively), from the beginning
of its outburst to 2010-10-05. Every point represents one RXTE observation and the solid line joins consecutive observations.
Intensity and hardness are defined in \S \ref{par:data_analysis}.
The black symbols mark the observations for which we analysed power density and energy spectra, and the black downward triangles 
mark the RXTE observations performed quasi simultaneously to Swift observations.
The numbers help the reader to follow to evolution of the outburst (starting from \#1).
The inset in the right panel shows the RID for the whole outburst.
Dashed lines mark 30\%, 20\% and 10\% fractional rms levels.}
\label{fig:HID_licu}
\end{figure*}

Swift J1753.5-0127 was discovered in the hard X-ray band with the Burst Alert Telescope (BAT) aboard Swift on 2005 May 30 (Palmer et al. 2005).
After the discovery, the source flux reached a peak of 200 mCrab on 2005 July 9, as observed by the All Sky Monitor (ASM) on board the Rossi
X-ray timing explorer (RXTE) satellite (see the light curve in Figure \ref{fig:HID_licu}).
Subsequently the flux started to decrease and then stalled at a level of $\sim$20 mCrab (2-20 keV) for more than $\sim$ 6 months.
This is an unusual behaviour for a transient, but even more unusual is the subsequent shallow flux rise which has been ongoing
since roughly June 2006, with a steepening on June 2008 (Krimm et al. 2008). Figure \ref{fig:HID_licu} shows that despite this
re-brightening, the source never came back to the flux level observed in the first months of its outburst.
The source has not returned to quiescence and at the moment of writing this paper it is still active (e.g. Soleri et al. 2012).
Although the X-ray spectral hardness of Swift J1753.5-0127 has not remained constant (Zhang et al. 2007, Ramadevi \& Seetha 2007,
Negoro et al. 2009), no state transition to the intermediate states or to the HSS was reported so far.
The mass of the compact object in the binary has not been dynamically measured. However, the
strongest hint that the system harbours a black hole comes from the hard power-law tail in the X/$\gamma$-ray energy
spectrum detected with INTEGRAL up to $\sim$600 keV (Cadolle Bel et al. 2007), as no neutron-star low-mass X-ray binary has ever been
detected above $\sim 200$ keV (see e.g. Falanga et al. 2007).
\begin{figure*}
\begin{tabular}{c}
\resizebox{15.5cm}{!}{\includegraphics{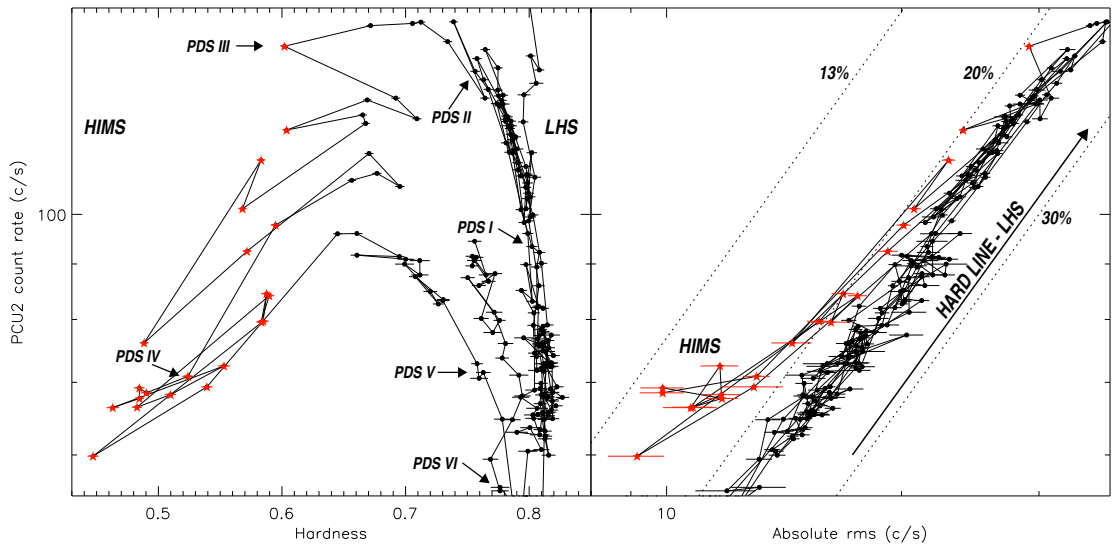}}
\end{tabular}
\caption{Zoom of the HID and the RID on the ``failed transition'' of Swift J1753.5-0127. Each point corresponds to an entire RXTE observation and the
solid line joins consecutive observations. The red stars represent the points with hardness $h < 0.63$. 
In the RID we drew the dotted lines at 30\%, 20\% and 13\% fractional rms levels and we also marked the hard line,
which identifies the LHS observations.
In the HID, we labelled the PDS that are shown in Figure \ref{fig:PDS}.}
\label{fig:RID_zoom}
\end{figure*}

Swift J1753.5-0127 is certainly interesting for a number of reasons. For instance, the system follows the lower track
in the X-ray/radio luminosity plane (Cadolle~Bel et al. 2007; Soleri et al. 2010), together with an increasing number of black holes
(Coriat et al. 2011, see also Corbel et al. 2012 for the most recent version of the X-ray/radio luminosity plane).

Secondly, Swift J1753.5-0127 would be the black hole with the second shortest orbital period, according to Zurita et al. (2008)
and Durant et al. (2009), who reported a $\sim$3.2 hr modulation in the optical light curves (with absorbing dips recurring every $\sim 2.4$ hr, the BHT
MAXI J1659-152 is the system with the shortest orbital period; Kuulkers et al. 2012).

In this paper we study the evolution of the outburst of Swift J1753-0127, from its beginning to
October 2010. Our data analysis is presented in \S \ref{par:data_analysis}.
In \S \ref{par:fund_diag} we describe the general features of the whole outburst, making use of the X-ray light curve, HID and RID;
then we focus on X-ray spectral and timing properties of the source during the softening, which took place between 
June 2009 and May 2010 (\S \ref{par:result_timing}, \S \ref{par:result_energy}). In the discussion (\S \ref{par:discussion}) we show that the
overall spectral and timing properties of the source are similar to those displayed by the majority of the BHTs in the hard state, and we bring further evidence
towards the identification of the accreting object as a black hole (\S \ref{par:nature}). In \S \ref{par:outburst} we discuss the peculiarities
of Swift J1753.5-0127, making a comparison with other accreting systems. Our conclusions are presented in \S \ref{par:conclusions}.

\section{Observations and data analysis} \label{par:data_analysis}
In this section we describe our analysis: data from the Proportional Counter Array (PCA) aboard RXTE were used for producing the light curves,
as well as for the variability studies; for the spectral analysis we used data from both RXTE and the Swift satellite.

\subsection{Light curve, HID and RID} \label{par:fund_diag_analysis}
We used 263 PCA observations performed until 2010-10-05 (Obs. ID 95105-01-35-01)
to produce the light curve and the HID in Figure \ref{fig:HID_licu}. The restriction to this interval is motivated by the fact that in October 2010, after
a soft excursion, the source is again steadily in the LHS, featuring similar properties to those displayed before the softening.
For each observation, we used the standard RXTE software (within HEASOFT V. 6.11) to obtain the background and
dead-time\footnote{for details on the dead-time correction see the RXTE Cook Book: http://heasarc.nasa.gov/docs/xte/recipes/pca\_deadtime.html}
corrected count rate from the Proportional Counter Unit 2 (PCU2) in Standard 2 (STD2) data mode (Jahoda et al. 2006), in the channel range 0-31 (2-15 keV).
We defined the spectral hardness $h$ as the ratio of counts in the STD2 channel bands 11-20 and 4-10 (6.1-10.2 keV and 3.3-6.1 keV, respectively).
Using the same method described in Mu{\~n}oz-Darias et al. (2011),
we produced the RID presented in the inset of Figure \ref{fig:HID_licu} and in Figure \ref{fig:RID_zoom}, calculating the fractional rms amplitude
in the frequency band 0.1-64 Hz, for the PCA absolute channels 0-35 (2-15 keV).
The absolute rms (on the x axis of the RID) was computed by multiplying the fractional rms by the net source count rate.
Using the same method, we also obtained the fractional rms using a soft and a hard band (channels 0-13 and 14-35; this corresponds to
2-6 keV and 6-15 keV, respectively), as well as the rms spectrum for observation 91094-01-01-04, which was produced by
computing the fractional rms in 7 energy bins in the 2-20 keV energy band.
For the rms spectrum, the specific background associated with each energy bin was considered, and the rms was computed following standard procedures
(see for instance Berger \& van der Klis 1994, Rodriguez \& Varni{\`e}re 2011).
Throughout the paper, we will refer to the fractional rms amplitude as ``rms'' or ``fractional rms''
while we will explicitly use the adjective ``absolute'' when referring to the absolute rms.

From the HID of Swift J1753.5-0127 it is apparent that the spectral hardness did not remain
constant during the whole outburst: the source softened between approximately 2009 July and 2010 May (\#5 to \#7 in
Figure \ref{fig:HID_licu}) and then returned to its initial hardness level in 2010 July (\#8 in Figure \ref{fig:HID_licu}).
Detailed timing and spectral analysis of this soft excursion is described in \S \ref{par:timing_analysis}
and \S \ref{par:spectral_analysis}. Despite the softening, Swift J1753.5-0127 never went to the HSS
(see \S \ref{par:discussion}): for this reason we will refer to this phase of the outburst as the ``failed transition''.

\subsection{Timing analysis} \label{par:timing_analysis}
We produced power density spectra for 67 RXTE observations performed between 2009-06-27 and 2010-07-28 (Obs. IDs 93105-02-36-00 and
95105-01-25-01, respectively; see the black points in Figure \ref{fig:HID_licu}). The restriction to this interval
is due to the fact that we want to study in detail the source behaviour during the ``failed transition'', by means of
timing and spectral analysis.
Power density spectra were computed (from data in GoodXenon mode; Jahoda et al. 2006) using standard fast Fourier transform techniques
(van der Klis 1989) on segments of 128 s accumulated
over all the PCA energy channels (0-249) for all the available PCUs, with a Nyquist frequency of 4096 Hz.
For every RXTE observation we calculated one average PDS, rebinned it logarithmically and also subtracted the Poissonian noise (Zhang et al. 1995).
If two contiguous observations were performed within 24 hours, we averaged their power spectra together.
The resulting PDS were normalized according to Leahy et al. (1983).
The fitting was carried out using the sum of Lorentzian components (normally one or two), as defined in Belloni, Psaltis \& van der Klis (2002a). 
Throughout the paper we will quote the characteristic frequency $\nu_{max}$ at which the Lorentzians attain their maximum in the
$\nu \, P_{\nu}$ representation, and the quality factor $Q$ (Belloni et al. 2002a), where $\nu_{max}=\nu_0(1+1/4Q^2)^{1/2}$ and $Q=\nu_0/2\Delta$
($\nu_0$ is the Lorentzian centroid frequency and $\Delta$ is its half-width-at-half-maximum).
The best-fitting parameters, as well as their $1\sigma$ errors (determined using $\delta \chi^2 = 1$), are reported in Tables \ref{tab:log_PDS_fit} and
\ref{tab:log_PDS_fit_bis} in the Appendix.
\begin{figure}
\begin{tabular}{c}
\resizebox{8.4cm}{!}{\includegraphics{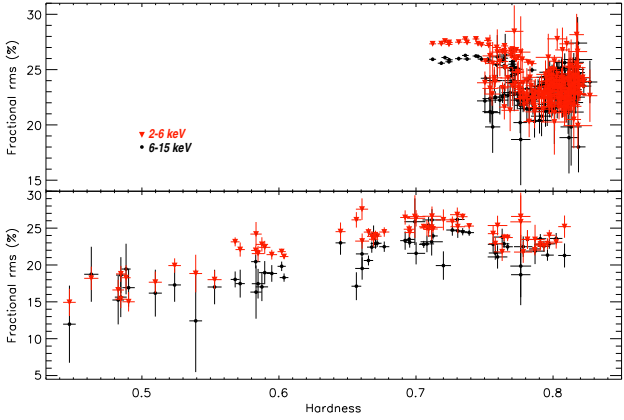}}
\end{tabular}
\caption{Hardness-rms diagram of Swift J1753.5-0127, using soft (2-6 keV, red-downward triangles) and hard (6-15 keV, black-filled circles)
energy bands for computing the rms. The lower panel shows the observations in the ``failed transition''
(black symbols in Figure \ref{fig:HID_licu}), the upper panel all the other observations (grey points in Figure \ref{fig:HID_licu}).
Note the different scale on the y axis of the 2 panels.}
\label{fig:hardness-rms_diag}
\end{figure}
  
\subsection{Spectral analysis} \label{par:spectral_analysis}
Here we describe the procedures to extract and fit energy spectra from RXTE and Swift data.
A complete log of the models used for the spectral fitting, as well as the most relevant fitting parameters, are reported in Tables
\ref{tab:log_fit_RXTE_spectra_1}, \ref{tab:log_fit_RXTE_spectra_2} and \ref{tab:log_fit_Swift_spectra} in the Appendix.

\subsubsection{RXTE}
We produced energy spectra from the same RXTE observations used for the timing analysis described in \S \ref{par:timing_analysis}. We applied standard
filters and extracted PCA spectra from STD2 data (PCU2 only, using all layers). The data were background and dead-time corrected. We generated response matrices
and, following Motta et al. (2011), we fitted the spectra in the energy range 4-40 keV.
In order to account for residual uncertainties in the instrument calibration we
applied $0.6 \%$ systematic error\footnote{See
http://www.universe.nasa.gov/xrays/programs/rxte/pca/doc/
rmf/pcarmf-11.7/\#head-27884cefb2a102a7e53547f1631cbeab44224a04
for a detailed discussion on the PCA calibration issues.}.
The energy range was reduced to 4-25 keV when the PCA spectra were fitted together with Swift/XRT spectra.
For the observations performed before December 2009, we also extracted background and dead-time corrected spectra from the
Cluster B of the High Energy Timing Experiment (HEXTE aboard RXTE) and we fitted them in the 20-200 keV energy range, after generating
response matrices. On December 2009 HEXTE Cluster B stopped rocking (during normal operation, the two clusters rock on
and off target to collect background data), hence we did not extract HEXTE spectra after that date.
Given the large Poissonian uncertainties in the HEXTE spectra due to the low number of counts, no systematic error was added.
\begin{figure}
\begin{tabular}{c}
\resizebox{8.3cm}{!}{\includegraphics{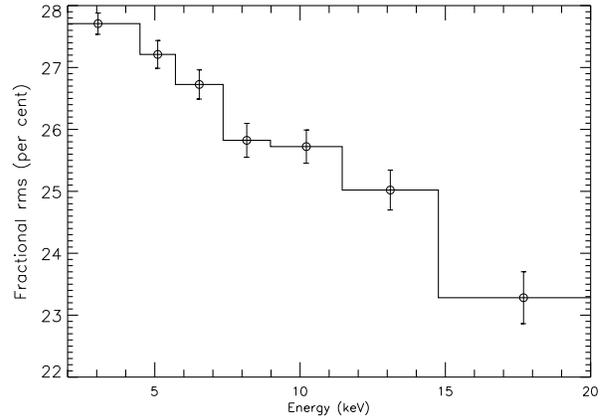}}
\end{tabular}
\caption{Rms spectrum of Swift J1753.5-0127 in the frequency band 0.1-64 Hz. This rms spectrum was obtained from 
the RXTE observation performed on 2005-07-06 (obs. ID 91094-01-01-04), at the beginning of the outburst.}
\label{fig:rms_spec}
\end{figure}

\subsubsection{Swift}   \label{par:spectral_Swift}
We produced energy spectra from all the seven pointed Swift observations performed during the softest phase of the outburst
(see Table \ref{tab:log_fit_Swift_spectra} in the Appendix). All the observations were performed quasi
simultaneously to RXTE (within 36 hours).
Two Swift observations (obs. IDs 00031232017 and 00031232018) were taken quasi simultaneously to the same RXTE pointing
(Obs. ID 95105-01-16-00) with a time delay of $\sim 24$ hours and less than $\sim 11$ hours, respectively.
We therefore fitted simultaneously the RXTE observation and the closest Swift observation in time.
We processed the XRT data using {\it xrtpipeline}
with standard quality cuts. Our data were all collected in windowed timing mode, in which a 1D image is obtained by adding the data along the
central 200 pixels in a single row (see Hill et al. 2004). We extracted source and background spectra using
box regions with length of $\sim 30$ and $\sim 15$ pixels, respectively. Exposure maps and ancillary response files were
generated using {\it xrtexpomap} and {\it xrtmkarf}. The latest response matrix files (v12) were taken from the calibration
database. We applied a systematic error of $2.5 \%$ (Campana et al. 2008)
and we grouped the spectra to a minimum of 20 counts per bin. The XRT spectra were fitted in the 0.8-8 keV energy range.
\begin{figure*}
\begin{tabular}{c}
\resizebox{17cm}{!}{\includegraphics{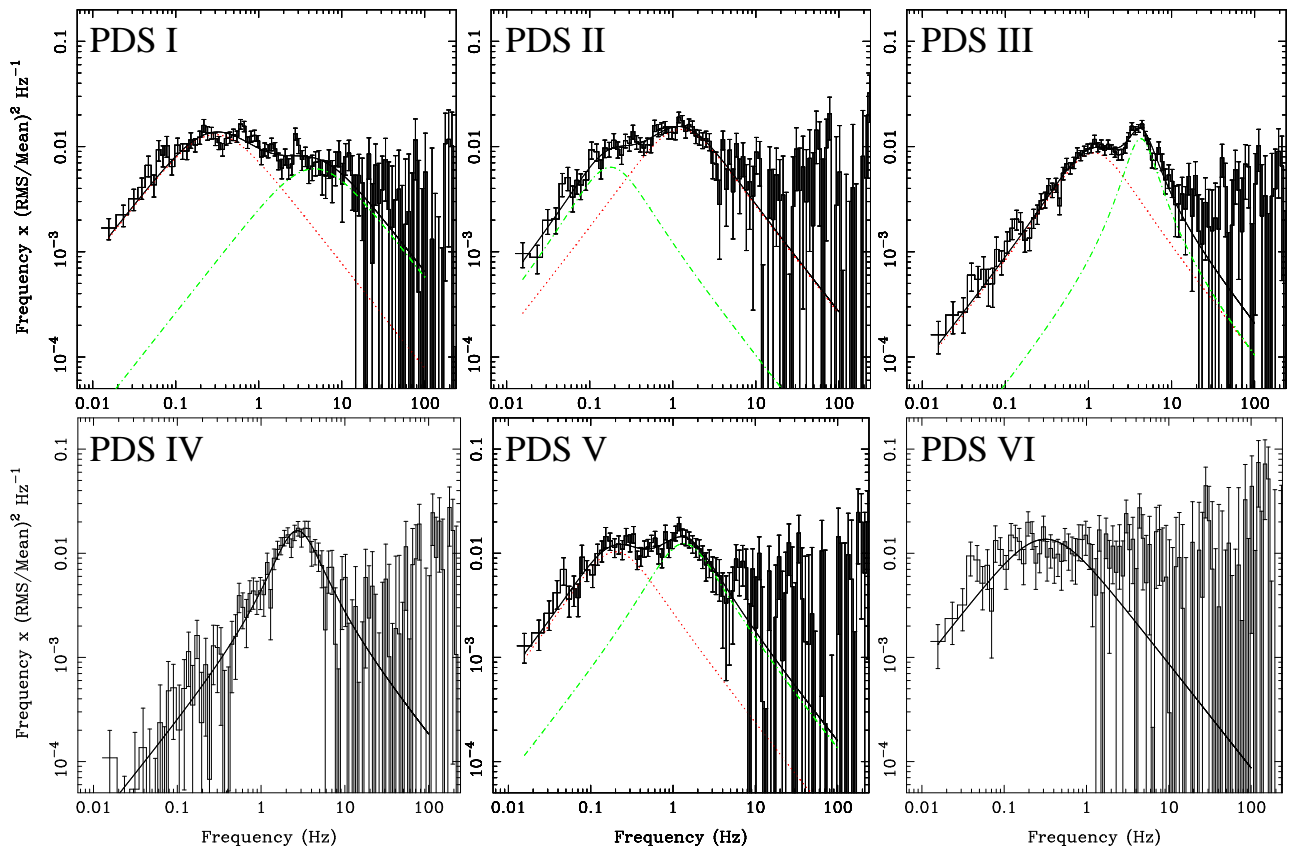}}
\end{tabular}
\caption{Fits to six power density spectra of Swift J1753.5-0127 extracted in the ``failed transition''.
We plotted all the Lorentzians, as well as the total fit function. PDS I to VI correspond to observations 93105-02-37-00, 93105-02-48-00,
93105-02-58-00, 95105-01-12-00, 95105-01-23-00 and 95105-01-25-00, respectively.
We labelled these observations in the HID in Figure \ref{fig:RID_zoom}.
In PDS I and VI we fixed the quality factor of the Lorentzians to $Q=0$.}
\label{fig:PDS}
\end{figure*}

\subsubsection{Fitting procedures}
We fitted the energy spectra using the standard \textsc{Xspec v11.3} fitting package (Arnaud 1996).
For simultaneous RXTE and Swift observations we combined the PCA+XRT spectra to fit them together.
In order to account for flux cross-calibration uncertainties, we multiplied the continuum model by a constant.
All model parameters, except these multiplicative constants, were linked between the different instruments.
When fitting the RXTE spectra alone, we fixed the interstellar absorption to the average column density value obtained by Hiemstra et al. 2009
($N_H = 1.7 \times 10^{21}$cm$^{-2}$) while we let it free to vary when including also Swift/XRT data. 
The RXTE spectra alone can be fitted using an absorbed broken (elbow-shaped) power law.
However, when fitting RXTE+Swift spectra, a disc blackbody at low energies and a Gaussian emission line were required
in order to obtain a statistically acceptable fit. 
The Gaussian component takes into account the escape peak of silicon at $1.9$ keV
(a known instrumental artefact\footnote{http://www.swift.ac.uk/xrtdigest.shtml\#res}).
Errors on the fit parameters were determined at $1 \sigma$ confidence level ($\delta \chi^2 = 1$).

\section{Results} \label{par:results}
In this section we first describe the general behaviour of Swift J1753.5-0127 during the whole outburst using the X-ray light curve,
HID and RID (\S \ref{par:fund_diag}), then we focus on the power and energy spectra of the source
in the ``failed transition'' (\S \ref{par:result_timing} and \S \ref{par:result_energy}).

\subsection{Outburst evolution} \label{par:fund_diag}
%
During the initial months of the outburst the source follows a hard line in the RID (a linear relation between the absolute rms and the count rate,
see the inset of Figure \ref{fig:HID_licu}), with fractional rms decreasing from $27 \%$ to $\sim 22$\% (in correspondence of \#2 and \#3 in the HID
in Figure \ref{fig:HID_licu}, respectively).
A clear softening took place in July 2009 (\#5, \#6, \#7), followed by a hardening at lower count rate (from \#7 to \#8). From \#8 to \#9
the source flux increased again, together with a weak softening (from $h \sim 0.78$ to $h \sim 0.75$).
Figure \ref{fig:RID_zoom} shows a zoom of the HID and the RID on the ``failed transition'' of Swift J1753.5-0127:
one can see that the points characterized by the lowest spectral hardness (marked by red stars; we arbitrarily put
a threshold at a hardness $h=0.63$) correspond to the observations with the lowest fractional rms. The RID in Figure \ref{fig:RID_zoom} also shows that
in the softest part of the ``failed transition'' the source does not follow the hard line.

In Figure \ref{fig:hardness-rms_diag} we plot the fractional rms amplitude in the soft and in the hard band (red triangles and black circles, respectively)
versus the hardness.
For clarity, we show the points from the ``failed transition'' in the bottom panel.
During the first weeks of the outburst (see the points at the lowest spectral hardness in the top panel of Figure \ref{fig:hardness-rms_diag}),
the rms features an inverted spectrum.
Following Gierli{\'n}ski \& Zdziarski (2005), we define inverted, flat and hard rms spectra the spectra in which the fractional
rms at low energies is higher, similar to or lower than the rms at high energies, respectively.
An example of inverted rms spectrum from the initial phases of the outburst is shown in Figure \ref{fig:rms_spec}.
Given the lower number of counts, for the rest of the outburst we infer the shape of the rms spectrum
from the values of the rms in the hard and in the soft bands in Figure \ref{fig:hardness-rms_diag}.
After the first weeks, the rms spectrum is consistent with being flat, including the observations from the
``failed transition'' in the bottom panel. However, a hard rms spectrum cannot be discarded for a handful of observations.
\begin{figure}
\begin{tabular}{c}
\resizebox{8.4cm}{!}{\includegraphics{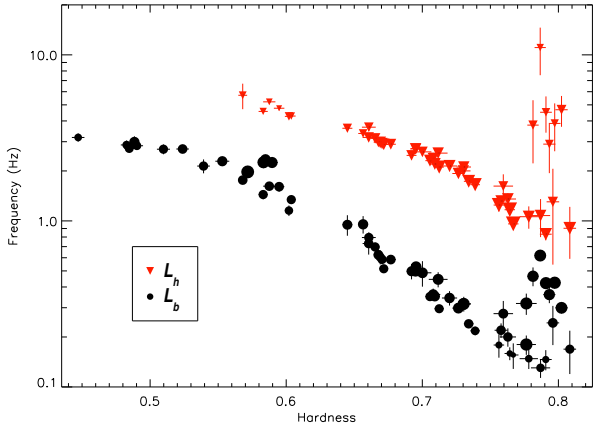}}
\end{tabular}
\caption{Frequency of the Lorentzians used to fit the power density spectra of Swift J1753.5-0127 (see Tables \ref{tab:log_PDS_fit}
and \ref{tab:log_PDS_fit_bis}) as a function of the spectral hardness.
The black circles and the red upward triangles represent $L_b$ and $L_h$, respectively.
The area of the symbols is proportional to the fractional rms of the peaks. The rms lies in the ranges $\sim 9-22$\% and $\sim 8-21$\%
respectively for $L_b$ and $L_h$.}
\label{fig:evoluz_QPO}
\end{figure}
\subsection{Power density spectra} \label{par:result_timing}
Figure \ref{fig:PDS} shows six power density spectra fitted with one or two Lorentzians, representative of the source behaviour
during the ``failed transition''.
Following van Straaten, van der Klis \& M{\'e}ndez (2003), if the $Q$-value of a Lorentzian
becomes negative in the fit, we fix it to zero.
\begin{figure}
\begin{tabular}{c}
\resizebox{8.4cm}{!}{\includegraphics{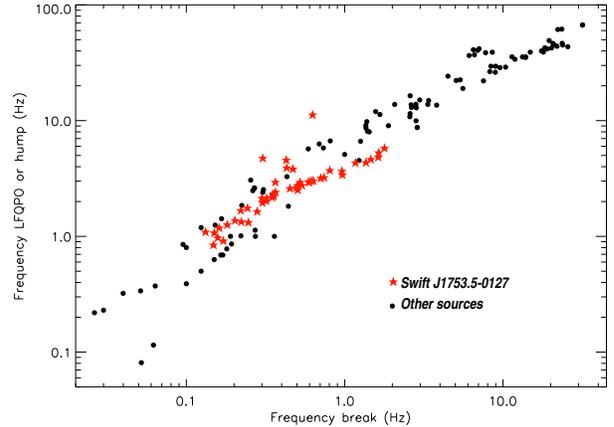}}
\end{tabular}
\caption{Frequency of the hump $\nu_h$ as a function of the break frequency $\nu_b$ for the power spectra of Swift J1753.5-0127 in which
both components are present (red stars).
The black circles represent the data points from Figure 11 of Belloni et al. (2002a), including only those observations in which
either $L_{LFQPO}$ or $L_h$ is present. The correlation between the frequencies $\nu_b$ and $\nu_{LFQPO}$ (or $\nu_h$)
is normally called WK relation (Wijnands \& van der Klis 1999).}
\label{fig:wk}
\end{figure}
In Figure \ref{fig:evoluz_QPO} we plot the frequency of the Lorentzians as a function of the hardness: two broad peaks can be
identified in the PDS (with quality factor $Q$ below $\sim 1$ and $\sim 2$, respectively), decreasing in frequency with the hardness.
However, in the range $0.75 < h < 0.81$, the frequency increases without being correlated with the hardness. Most of the
power spectra obtained from the observations in this hardness range are fitted with Lorentzians in which we fixed the value of $Q$ to zero.
Although the first peak, represented by black circles, is present in the whole hardness range, the second peak
(downward red triangles) disappears at $h < 0.57$. PDS IV in Figure \ref{fig:PDS} represents an example of a power spectrum
from an observation characterized by low $h$ and fitted with one Lorentzian.
We found no significant correlation between the values of the fractional rms
(that is proportional to the area of the symbol) and the hardness for either of the two Lorentzians.

Figure \ref{fig:wk} shows the frequency of the low-frequency QPO ($\nu_{LFQPO}$) or of the hump ($\nu_h$) for a number of black holes
and neutron stars in the hard state (commonly called WK correlation; Wijnands \& van der Klis 1999)
as a function of the break frequency $\nu_b$ (black circles).
We adapted the plot from Figure 11 of Belloni et al. (2002a), including only those observations in which either $L_{LFQPO}$
or $L_h$ is present, which is the most relevant case to our analysis. The hump ($L_h$) usually appears at frequencies similar to
the LFQPO ($L_{LFQPO}$) but it is characterized by a lower quality factor ($Q < 2$). 
The plot also includes the data points for Swift J1753.5-0127 (red stars), obtained using the frequencies of the
two peaks described above. As one can see, the path of Swift J1753.5-0127 in the diagram overlaps to the one of the other
systems, despite its (possibly) shallower slope and two points which fall above the WK correlation. The higher frequency peak that we fit in the PDS of
Swift J1753.5-0127 (red downward triangles in Figure \ref{fig:evoluz_QPO}) is broad (with $Q < 2$, except for two cases,
see Table \ref{tab:log_PDS_fit}), and so we cannot identify it as the LFQPO commonly detected in the hard states.
Following Belloni et al. (2002a), we will refer to the lower and the higher frequency broad peaks in the power spectra of Swift J1753.5-0127
as $L_b$ (break) and $L_h$ (hump), respectively.
\begin{figure*}
\begin{tabular}{c}
\resizebox{15.5cm}{!}{\includegraphics{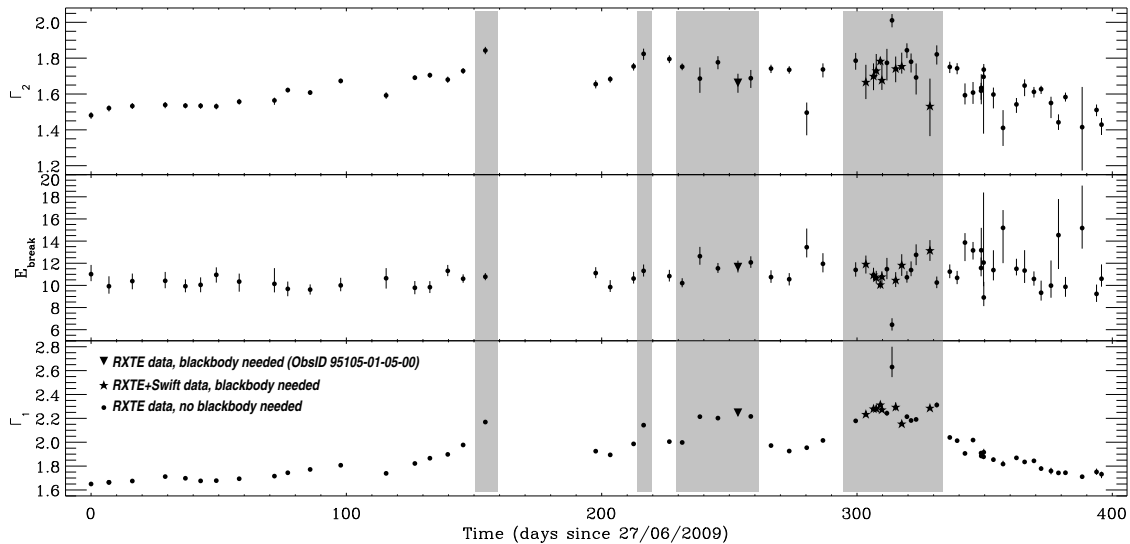}}
\end{tabular}
\caption{Evolution of the three parameters of the broken power-law used to fit the energy spectra of Swift J1753.5-0127 in Tables
\ref{tab:log_fit_RXTE_spectra_1}, Tables \ref{tab:log_fit_RXTE_spectra_2} and \ref{tab:log_fit_Swift_spectra}. The grey areas
mark the red stars in Figure \ref{fig:RID_zoom}.}
\label{fig:param_energy_spectra}
\end{figure*}

\subsection{Energy spectra}  \label{par:result_energy}
\begin{figure}
\begin{tabular}{c}
\resizebox{8.2cm}{!}{\includegraphics{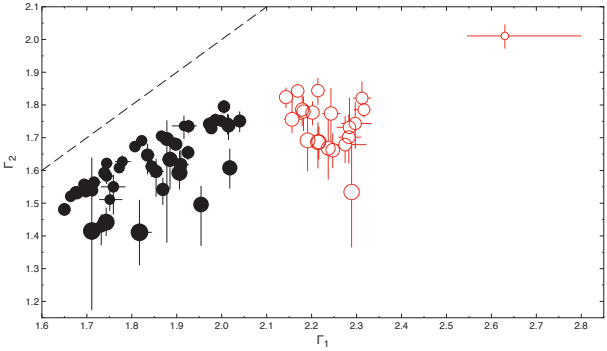}}
\end{tabular}
\caption{
Correlation between the photon indices of the broken power law component at energies below and above the break
($\Gamma_1$ and $\Gamma_2$, respectively), for all the spectral fits of Swift J1753.5-0127 in Figure \ref{fig:param_energy_spectra}. The area of the points is
proportional to the value of the energy break $E_{break}$. The dashed line corresponds to $\Gamma_1 = \Gamma_2$.
Red empty circles correspond to the red stars in Figure \ref{fig:RID_zoom} while the black filled circles mark the other observations in the
``failed transition''.}
\label{fig:Gamma2_VS_Gamma1}
\end{figure}

In Figure \ref{fig:param_energy_spectra} we report the evolution of the parameters of the broken power-law for all the spectral fits, where
$\Gamma_1$ and $\Gamma_2$ are the power-law photon indices at energies below and above the break, respectively. As expected, the observations in
the softest part of the HID (the red stars in Figure \ref{fig:RID_zoom}) correspond to the highest values of the photon
index $\Gamma_1$ (grey shaded areas in Figure \ref{fig:param_energy_spectra}).
From Figure \ref{fig:Gamma2_VS_Gamma1} we can see that $\Gamma_2$ and $\Gamma_1$ could be correlated (despite a large scatter):
a Spearman rank correlation coefficient indicates that a weak correlation probably exists, with correlation coefficient $\rho = 0.75$
and a low probability for the null hypothesis ($\lesssim 10^{-6}$).
The slope of the correlation becomes shallower at $\Gamma_1 \gtrsim 2$.
In this diagram, the area of the points is proportional to the value of $E_{break}$, suggesting that the largest values of $E_{break}$
are associated with the lowest $\Gamma_2$.

In our spectral fits we used a {\it diskbb} component six times
(in five spectra from PCA+XRT data and in one spectrum from PCA data) with temperature
in the range $0.30-0.42$ keV. This is slightly lower than the values usually observed in other BHTs in these states
(see e.g. Belloni et al. 2011).
Although an F-test suggests that in all these spectra the disc component significantly improves the fit\footnote{Formally, an F-test might not be
applicable here, since the function that represents the blackbody is limited (it must be positive). Nevertheless we use it to have
an indication of the improvement of the fits (see Protassov et al. 2002 for the applicability of the F-test).}, in four spectra the disc normalization cannot be
measured accurately (the ratio between the disc normalization and its negative error at 1 $\sigma$ is smaller than 3).
In order to understand whether the thermal component is really needed, we calculated the disc fluxes (or the 95\% upper limits
for the spectra in which the disc normalization is not well constrained), showing that all the values are
consistent within the errors. This suggests that, although our ability of constraining the parameters of the {\it diskbb} varied,
the disc itself did not vary much between these observations.
%

\section{Discussion} \label{par:discussion}
Despite several years of observations at different wavelengths, it is difficult to explain the evolution of the particularly
long outburst of Swift J1753.5-0127 (still ongoing at the moment of writing this manuscript) in the same framework as the majority
of the BHTs. Our observing campaign with RXTE and Swift provided important information about its evolution,
showing that the source did not go to the soft states and allowing us to make comparisons with other (classes of) sources.

As expected for a BHT in the LHS, Swift J1753.5-0127 draws a hard line in the RID and it features, at the beginning of the outburst
(when $h \lesssim 0.78$ and the count rate is above $\sim 160$ c/s/PCU2), its highest fractional rms (between $24 \%$ and $27 \%$).
Unlike the majority of the BHTs in the LHS (Belloni et al. 2011), the fractional rms
anticorrelates with the spectral hardness $h$ and correlates with the count rate (see Figure \ref{fig:HID_licu}).
This shows that, at least in this case, the value of $h$ and the position in the HID was not an
indication of an approaching spectral transition to the intermediate states.
In GX 339-4, Mu{\~n}oz-Darias et al. (2011) associated the transition from the LHS to the HIMS with a clear variation in the X-ray RID,
namely the source leaves the hard line and moves to the left-hand side of the diagram along another track. Interestingly,
this change in the RID is related to a sudden change in the X-ray/infrared correlation: in the 2002 outburst of GX 339-4,
the infrared flux (in the H band) dropped by a factor $\sim 17$ in about 10 days while in the same time interval the X-ray
flux only decreased by a factor $\sim 1.4$ (Homan et al. 2005).
Figure \ref{fig:RID_zoom} shows that the observations of Swift J1753.5-0127 with the lowest spectral hardness are also
characterized by the lowest fractional rms and they all lie on the left-hand side of the hard line.
Unlike GX 339-4, these observations do not follow a precise path in the RID but we observe several hardenings and
softenings. Nevertheless, we identify the observations with the lowest hardness and the lowest fractional rms as 
HIMS observations.
It is also worth pointing out that in Swift J1753.5-0127, differently from GX 339-4, it is not possible 
to mark the exact moment of the transition from the LHS to the HIMS, since we neither see a clear turn in the RID nor we
have simultaneous observations at longer wavelengths. It would be interesting to test whether the different behaviour in the RID is
related to the fact that Swift J1753.5-0127 and GX 339-4 follow the lower and the upper track in the
X-ray/radio domain (respectively, see Coriat et al. 2011). A comparison with other BHTs
is essential to test this possibility.
After this excursion in the HIMS, Swift J1753.5-0127 returned to the LHS, without passing through the soft
states (either SIMS or HSS).
\begin{table*}
\centering
\caption{List of RXTE observations used to compare the power spectra of black holes and neutron stars. The letters in the first column
identify the power spectra in Figure \ref{fig:comparison_PDS}. We extracted all the power spectra following the procedure
presented in \S \ref{par:timing_analysis}, with the only difference being that intervals of 256 s were used to accumulate the power spectra
of 4U 1705-44 and 4U 1608-52. The Table also shows the values of the $\chi^2$ and the degrees of freedom for the best fit, the break
frequency $\nu_b$, the rms in the frequency intervals $\nu_b - 100 \nu_b$ and $100\nu_b - 1000 \nu_b$ and their ratio.}
\label{tab:log_comparison}
\begin{tabular}{c c c c c c c c}
\hline
         &       &                &                  &                &  \multicolumn{2}{c}{rms (\%)}                   &           \\
PDS & Obs. ID & Fit function   & $\chi^2 /$d.o.f. & $\nu_{b}$ (Hz) & $\nu_b - 100 \nu_b$ & $100\nu_b - 1000 \nu_b$ & ratio rms \\ 
\hline
                                    \multicolumn{8}{c}{{\bf Black holes}}                                  \\       
                                    \multicolumn{8}{c}{{\it Swift J1753.5-0127}}                                     \\       
a) & 93105-02-52-00 & 2 Lorentzians & 175/116 & $0.16\pm0.01$ & $21.52\pm0.21$ & $7.82\pm1.03$ & $2.75\pm0.36$            \\
                                    \multicolumn{8}{c}{{\it XTE J1550-564}}                                          \\       
b) & 50137-02-05-01 & 7 Lorentzians & 511/414 & $0.15\pm0.02$ & $30.38\pm0.06$ & $10.41\pm0.14$ & $2.92\pm0.04$           \\
\hline
                                    \multicolumn{8}{c}{{\bf Neutron stars}}                                          \\ 
                                    \multicolumn{8}{c}{{\it 4U 1705-44}}                                             \\
c) & 40034-01-04-05 & 3 Lorentzians & 85/97   & $0.22\pm0.02$ & $23.36\pm0.25$ & $14.27\pm0.93$ & $1.64\pm0.11$           \\
                                    \multicolumn{8}{c}{{\it 4U 1608-52}}                                             \\
d) & 93408-01-02-02 & 3 Lorentzians & 100/98  & $0.14\pm0.02$ & $16.77\pm0.33$ & $13.47\pm1.11$ & $1.25\pm0.11$           \\
e) & 93408-01-08-02 & 3 Lorentzians & 74/98   & $0.18\pm0.03$ & $18.16\pm0.25$ & $12.63\pm0.92$ & $1.44\pm0.11$           \\
f) & 93408-01-25-01 & 1 broken power law & 81/102 & $0.16\pm0.05$ & $19.72\pm0.41$ & $14.57\pm1.38$ & $1.35\pm0.13$       \\
\hline
\end{tabular}
\end{table*}

In \S \ref{par:result_timing} we showed that the frequencies of the Lorentzians used to fit the power density spectra of Swift J1753.5-0127
follow the WK relation, as other BHTs in the LHS and HIMS.
During the ``failed transition'' we do not observe any LFQPO.
A LFQPO (commonly called ``type C''; see Belloni et al. 2011 and references therein) is often detected in the PDS of black holes in the 
LHS and it is considered a signature of the HIMS. An example of another source which does not feature the type-C QPO in the PDS is the black
hole Cyg X-1 (Pottschmidt et al. 2003). 
It is interesting to note that Zhang et al. (2007) and Ramadevi \& Seetha (2007) detected a LFQPO
in the power spectra of Swift J1753.5-0127 from the first months of the outburst. The properties of this LFQPO are consistent with a type C, although
both Zhang et al. (2007) and Ramadevi \& Seetha (2007) do not use this nomenclature to classify the QPO.

During the initial weeks of the outburst the rms spectrum is inverted, without any indication of flattening at low energies (see the rms spectrum
in Figure \ref{fig:rms_spec}). Although the majority of the BHTs in the LHS features flat spectra, inverted spectra have already been observed
(e.g. in XTE J1650-500, see Figure 5 in Gierli{\'n}ski \& Zdziarski 2005). Zdziarski et al. (2002) and Gierli{\'n}ski \& Zdziarski (2005) explained
this type of spectrum by considering a seed photon input to the Comptonizing corona with variable energy, that causes a pivoting
of the energy spectrum around $\sim 20-50$ keV and a decline in fractional rms above $\sim 5$ keV. In XTE J1650-500, the absence of a flattening
suggests that the disc from which the seed photons originates is very cold (a disc with $kT \sim 0.7$ keV would give a flattening
in the rms spectrum below $\sim 2$ keV, Gierli{\'n}ski \& Zdziarski 2005).
In \S \ref{par:result_energy} we showed that an absorbed broken power law is sufficient to fit all the spectra from the observations in the
``failed transition'', except for the softest observations where a cold disc is needed to fit the spectrum.
Chiang et al. (2010) analysed Swift and RXTE observations of Swift J1753.5-0127 performed when the source was in the LHS,
between July 2005 and July 2007. They found that a thermal component is needed to fit the energy spectra,
with temperature below $\sim 0.25$ keV. These results confirm that the disc is indeed too cold to produce a flat rms spectrum at
low energies.
After the first weeks the rms spectrum changes and becomes, for the rest of the outburst, consistent with being either
flat or slightly inverted.
Flat rms spectra can be easily explained by assuming that the entire energy spectrum (dominated by a
Comptonizing component) varies in normalization (luminosity) but not in spectral shape
(except possibly the high energy cut-off; Gierli{\'n}ski \& Zdziarski 2005).
\begin{figure*}
\begin{tabular}{c}
\resizebox{17cm}{!}{\includegraphics{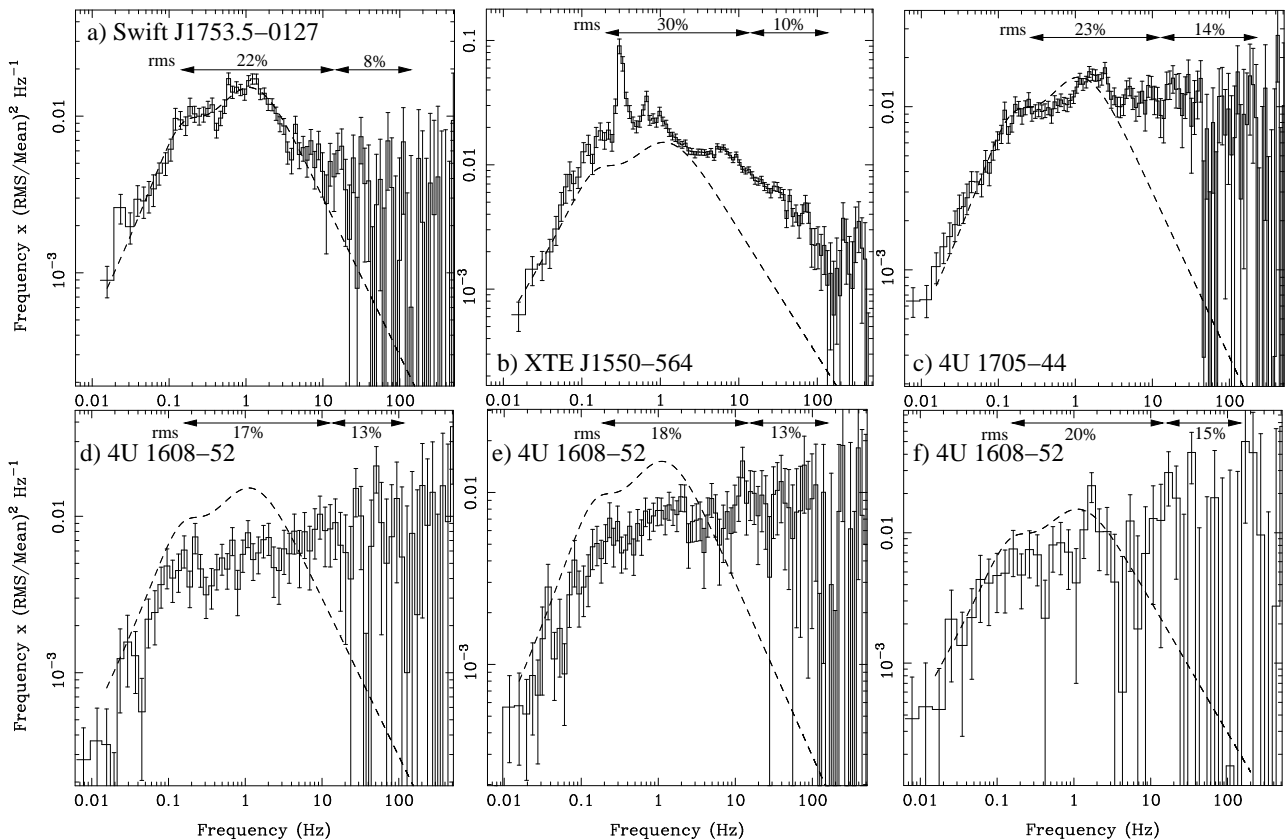}}
\end{tabular}
\caption{Power density spectra extracted from the RXTE observations of Swift J1753.5-0127, XTE J1550-564, 4U 1705-44 and 4U 1608-52
listed in Table \ref{tab:log_comparison}. In all the panels, the dashed line represents the best fitting function to the power spectrum
of Swift J1753.5-0127 reported in panel a). For each power spectrum we show the values of the fractional rms computed in the frequency intervals
$\nu_b - 100 \nu_b$ and $100\nu_b - 1000 \nu_b$ (see Table \ref{tab:log_comparison}).
The scale on the y axis is not the same for all the PDS.}
\label{fig:comparison_PDS}
\end{figure*}

\subsection{The nature of the compact object} \label{par:nature}
Although our results are consistent with Swift J1753.5-0127 being a BHT in the hard states, the nature of the compact object
(black hole or neutron star) still needs to be proved.
The weak correlation between the photon indices of the broken power law used to fit the energy spectra (see Figure \ref{fig:Gamma2_VS_Gamma1})
suggests that the spectral shape of the source evolves smoothly with the hardness between the LHS and the HIMS.
Interestingly, a similar correlation (with higher Spearman rank correlation coefficient $\rho = 0.92$) has already been reported by Wilms
et al. (2006) in the dynamically confirmed black hole Cyg X-1 (Herrero et al. 1995). Similarly to Swift J1753.5-0127 in the ``failed transition'',
Cyg X-1 goes through frequent spectral hardenings and softenings.
Moreover, both Cyg X-1 and Swift J1753.5-0127 (only in the ``failed transition'') lack the type-C QPO, which is usually observed
in the PDS of many black holes in the hard states.
This similarities between the two systems do not constitute a proof that Swift J1753.5-0127 contains a black hole, nevertheless
they can be considered another hint in this direction.

Sunyaev \& Revnivtsev (2000) analysed a sample of 9 black holes and 9 neutron stars in the hard state and noted that the black holes show
a steeper decline in the power spectra at frequencies higher than 10-50 Hz than the neutron stars.
%
Here we compare one power spectrum of Swift J1753.5-0127 in the LHS with power spectra from a dynamically confirmed BHT (XTE J1550-564;
see e.g. Orosz et al. 2011) and two secure neutron star atoll sources (4U 1705-44 and 4U 1608-52) in the hard (extreme island)
state (see van der Klis 2006 and references therein).
The purpose of our exercise is to obtain more information about the compact object in Swift J1753.5-0127 and not to perform
a systematic study of the differences between the power spectra of black holes and neutron stars.
In order to have a criterion to compare PDS from different sources, we only considered power spectra with similar break
frequency $\nu_b$. For each power spectrum, we calculated the fractional rms in the frequency intervals $\nu_b - 100 \nu_b$
and $100\nu_b - 1000 \nu_b$. The details of this analysis, as well as the power spectra,
are reported in Table \ref{tab:log_comparison} and in Figure \ref{fig:comparison_PDS}, respectively.
\begin{figure*}
\begin{tabular}{c}
\resizebox{17cm}{!}{\includegraphics{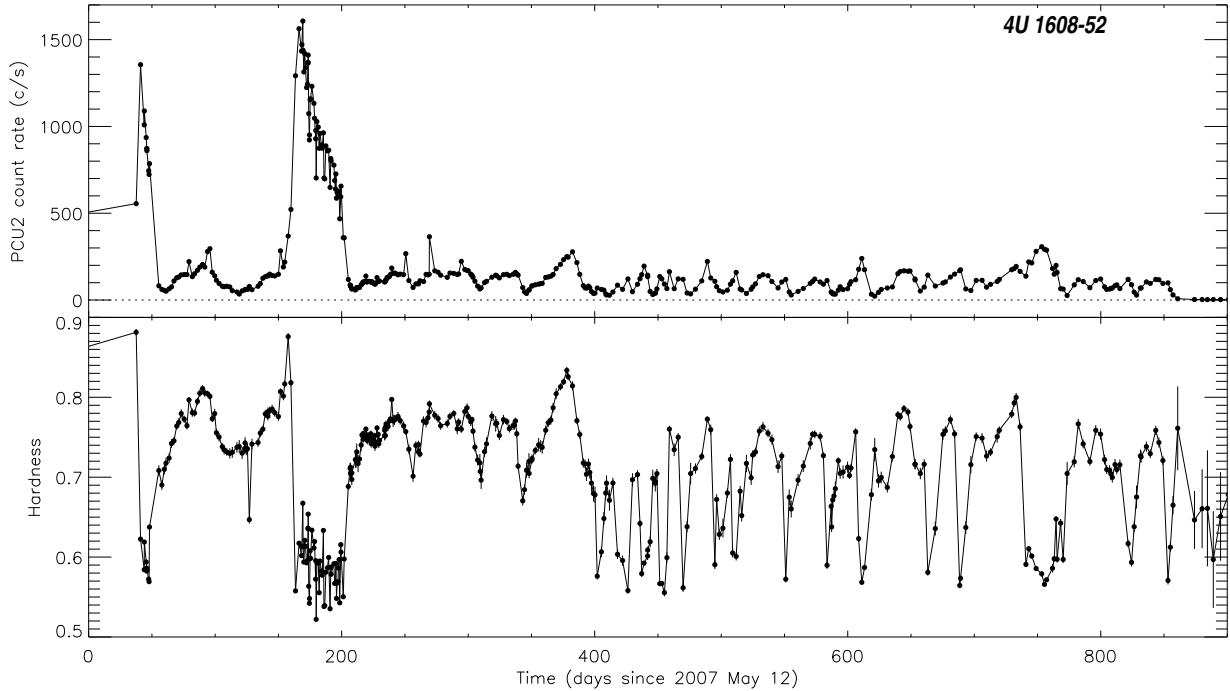}}
\end{tabular}
\caption{RXTE/PCA light curve and hardness curve of the neutron star 4U 1608-52 (top and bottom panels, respectively),
from 2007 May 12. Every point represent one RXTE observation. We obtained the count rate in the STD2 channel range 4-45 (3.3-20.6 keV) while
the hardness is the ratio between the count rate in the STD2 channel intervals 11-20 and 4-10 (6.1-10.2 keV and 3.3-6.1 keV, respectively).}
\label{fig:4U1608}
\end{figure*}
In our sample, neutron stars have a smaller drop (the ratio between the fractional rms in the
frequency intervals $\nu_b - 100 \nu_b$ and $100\nu_b - 1000 \nu_b$) in rms at high frequencies than black holes: a drop by a factor of $1.3-1.6$ in neutron
stars and by a factor $2.9$ in the BHT XTE J1550-564, whereas Swift J1753.5-0127 shows a $2.8$ factor drop.
The difference is already evident inspecting Figure \ref{fig:comparison_PDS}, in which the power spectra are overplotted to the best
fitting function to the spectrum of Swift J1753.5-0127.
This result is a further hint that Swift J1753.5-0127 hosts a black hole.

\subsection{Swift J1753.5-0127: a peculiar outburst} \label{par:outburst}
An interesting aspect of the outburst of Swift J1753.5-0127 concerns the morphology of the X-ray light curve, namely the long period at low
and slowly increasing flux that took place after the end of the initial flare and lasted more than two years.
Although such light curves are uncommon, a similar behaviour has already been observed in other X-ray binaries, for instance in the neutron star
4U 1608-52. This source usually undergoes regular outbursts, spaced out by periods of quiescence in which the PCA count rate is consistent with zero.
Figure \ref{fig:4U1608} shows the PCA light curve and hardness curve of 4U 1608-52 for a period of $\sim 900$ days from 2007 May 12,
where  we observe two outbursts.
In between the two outbursts and after the second one, the system goes into prolonged low-flux phases (with count rate below $\sim
400$ c/s/PCU2) in which it features several mini-outbursts, characterized by short duration (tens of days) and a high hardness level
(above $\sim 0.7$). The light curves of Swift J1753.5-0127 and 4U 1608-52 (focussing on Figure \ref{fig:4U1608})
presents some similarities: both sources had an initial flare (Swift J1753.5-0127 in 2005; 4U 1608-52 between days $\sim 150$ and $\sim 210$ in
the light curve in Figure \ref{fig:4U1608}) that was not
followed by a return to quiescence. In both cases, the systems neither went back to the flux levels observed at the peak of the initial flares
nor had a transition to the soft states.

It is interesting to note that the light curves of Swift J1753.5-0127 and 4U 1608-52 are reminiscent of those observed in a class
of cataclysmic variables, the so-called Z Camelopardalis (Z Cam).
Z Cam undergo regular outbursts like other subgroups of dwarf novae, followed by periods (weeks to months long) of standstills,
in which their optical luminosity remains constant at a level close to, but slightly below,
the peak of the outburst (see King \& Cannizzo 1998 for more details). Haswell \& King (2001) suggested that neutron star X-ray
binaries with short orbital period ($\lesssim 10$ hours, as the majority of the cataclysmic variables) could also feature standstills,
given their high mass transfer.
Considering the short orbital period of Swift J1753.5-0127 ($\sim 3.2$ hours; Zurita et al. 2008), it is possible that the equivalent of a standstill
followed its 2005 flare.
Although the orbital period of 4U 1608-52 is longer
($\sim 13$ hours; Wachter et al. 2002), its prolonged low-flux phase might be also interpreted as a standstill, possibly favoured
by the irradiation from the neutron star.

\section{Conclusions} \label{par:conclusions}
We observed the black-hole candidate Swift J1753.5-0127 with RXTE for more than five years. We studied the general evolution
of the outburst (still ongoing at the moment of writing this manuscript) making use of the X-ray light curve,
rms-intensity diagram and the hardness-intensity diagram. The source spent most of the time in the hard state and,
more than 4 years after the beginning of its outburst, moved to the hard-intermediate state. We followed the state transition
using power density and energy spectra (from RXTE and the Swift satellite) and showed
that the system returned to the hard state after a few months, without passing through the soft states.

During this ``failed transition'' Swift J1753.5-0127 features properties which are reminiscent of those observed in
the dynamically confirmed black hole Cyg X-1. First of all, both sources show frequent softening and hardening on
timescales of days, without leaving the hard states. Interestingly, the type-C QPO is not detected in Swift J1753.5-0127 during the
``failed transition'', although it is usually considered a signature of black holes in the hard states. Finally, in the energy spectra
of Swift J1753.5-0127 we observe a (weak) correlation between the photon indices of the broken power law similar to that previously
reported for Cyg X-1.

Looking for further clues about the nature of the accretor in Swift J1753.5-0127 (black hole of neutron star), we showed that
in one power spectrum the drop in rms at high frequencies (above 100 times the break frequency) is higher than in
two neutron star systems and similar to the dynamically confirmed black hole XTE J1550-564, indeed suggesting that
Swift J1753.5-0127 hosts a black hole.

We also studied the peculiar evolution of the outburst, focussing on the prolonged period at low flux that followed the
initial flare. Although such phases are rare in low-mass X-ray binaries, they are not unobserved, as for instance in the
neutron star 4U 1608-52. We suggest that these low-flux intervals are equivalent to the standstills observed
in some cataclysmic variables, namely in the Z Camelopardalis. This can be possibly explained considering the short orbital
period of Swift J1753.5-0127 ($\sim 3.2$ hours), close to the typical orbital periods of cataclysmic variables.

\section*{Acknowledgments}
We thank Andrew King, Andrea Sanna and Chris Done for very useful discussion and suggestions.
PS acknowledges support from NWO (Netherlands Foundation for Scientific Research).
The authors also thank the Principal Investigator of the Swift mission, Neil Gehrels,
for approving and scheduling the Swift ToO (Target of Opportunity).
TMD acknowledges funding via an EU Marie Curie Intra-European Fellowship under contract no. 2011-301355.
Partially funded by the Spanish MEC under the Consolider-Ingenio 2010 Program grant CSD2006-00070:
‘First Science with the GTC’ (http://www.iac.es/consolider-ingenio-gtc/). SM and TB
acknowledge support from grant INAF-ASI I/009/10/0.
The research leading to these results has received funding from the European Community's Seventh Framework Programme
(FP7/2007-2013) under grant agreement number ITN 215212 \textquotedblleft Black Hole Universe\textquotedblright. 
PC acknowledges funding via an EU Marie Curie Intra-European Fellowship under contract no. 2009-237722.
ML acknowledges support from a NWO Rubicon fellowship.
This research has made use of data obtained through the High Energy Astrophysics Science Archive Research Center Online Service, provided by the NASA
(National Aeronautics and Space Administration) Goddard space flight center.



\appendix
\section[]{Timing and spectral analysis} \label{app_tables}
Here we report five tables in which we show the best fitting parameters of the power density spectra as well as the energy
spectra for the observations performed during the ``failed transition''.

\begin{table*}
\centering
\caption{Fit parameters of the power density spectra extracted from the 67 RXTE observations of Swift J1753.5-0127 marked by black symbols in
Figure \ref{fig:HID_licu}. The details on the extraction of the spectra and on the fitting procedures are described in
\S \ref{par:timing_analysis}. For each observation, the table shows the frequency $\nu_{max}$,
the quality factor $Q$ and the rms of $L_b$ and $L_h$. The values of the $\chi^2$ and the degrees of
freedom are also reported. This Table continues in Table \ref{tab:log_PDS_fit_bis}.}
\label{tab:log_PDS_fit}
\begin{tabular}{c c c c c c c c c}
\hline
           &                & \multicolumn{3}{c}{{\bf $L_b$}} & \multicolumn{3}{c}{{\bf $L_h$}} &                            \\
Obs. ID    & Date           & $\nu_b$ (Hz)  & $Q_b$ & rms$_b$ ($\%$)     & $\nu_h$ (Hz)   & $Q_h$ & rms$_h$ ($\%$)     & $\chi^2 /$d.o.f. \\
\hline
93105-02-36-00 & 2009-06-27 & $0.17\pm0.05$ &  0  & $15.68\pm2.40$ & $0.90\pm0.31$  & 0   & $18.01\pm1.88$ & 145/118          \\  
93105-02-37-00 & 2009-07-04 & $0.30\pm0.02$ &  0  & $20.18\pm0.40$ & $4.67\pm0.98$  & 0   & $13.90\pm0.56$ & 120/118          \\
93105-02-38-00 & 2009-07-13 & $0.36\pm0.04$ &  0  & $19.24\pm0.79$ & $2.90\pm0.96$  & 0   & $12.36\pm1.03$ & 130/118          \\  
93105-02-40-00 & 2009-07-26 & $0.62\pm0.04$ &  0  & $20.42\pm0.33$ & $11.07\pm3.52$ & 0   & $10.47\pm0.79$ & 121/115 \\
93105-02-41-00 & 2009-08-03 & $0.42\pm0.03$ &  0  & $20.79\pm0.45$ & $4.51\pm1.11$  & 0   & $12.02\pm0.67$ & 116/118          \\
93105-02-42-00 & 2009-08-09 & $0.24\pm0.06$ &  0  & $16.74\pm2.06$ & $1.30\pm0.76$  & 0   & $14.51\pm2.15$ & 159/118          \\
93105-02-43-00 & 2009-08-15 & $0.42\pm0.03$ &  0  & $20.87\pm0.55$ & $3.86\pm1.24$  & 0   & $11.64\pm0.84$ & 164/118          \\
93105-02-44-00 & 2009-08-24 & $0.15\pm0.02$ & $0.40\pm0.16$ & $12.40\pm1.59$ & $0.83\pm0.07$ & $0.25\pm0.15$ & $17.55\pm1.57$ &105/114\\
93105-02-46-00 & 2009-09-07 & $0.46\pm0.06$ &  0  & $19.43\pm1.11$ & $3.78\pm1.56$  & 0   & $13.15\pm1.34$ & 120/118          \\ 
93105-02-47-00 & 2009-09-12 & $0.16\pm0.03$ & $0.34\pm0.17$ & $9.36\pm1.12$  & $0.96\pm0.06$ &  0            & $20.93\pm0.48$ & 166/117 \\
93105-02-48-00 & 2009-09-21 & $0.18\pm0.03$ & $0.33\pm0.13$ & $12.13\pm1.49$ & $1.25\pm0.11$ & $0.16\pm0.11$ & $20.05\pm1.07$ & 127/116 \\
93105-02-50-00 & 2009-10-03 & $0.22\pm0.01$ & $0.33\pm0.06$ & $14.95\pm0.50$ & $1.65\pm0.05$ & $0.38\pm0.05$ & $18.71\pm0.42$ & 132/116 \\
93105-02-52-00 & 2009-10-21 & $0.16\pm0.01$ & $0.42\pm0.14$ & $11.17\pm0.98$ & $1.17\pm0.06$ & $0.08\pm0.07$ & $20.66\pm0.60$ & 175/116 \\
93105-02-54-00 & 2009-11-01 & $0.24\pm0.01$ & $0.37\pm0.07$ & $16.18\pm0.50$ & $1.73\pm0.05$ & $0.51\pm0.07$ & $18.16\pm0.60$ & 110/113 \\
93105-02-55-00 & 2009-11-07 & $0.30\pm0.01$ & $0.37\pm0.06$ & $15.65\pm0.45$ & $2.11\pm0.05$ & $0.49\pm0.05$ & $18.64\pm0.42$ & 133/116 \\ 
93105-02-56-00 & 2009-11-14 & $0.35\pm0.02$ & $0.30\pm0.05$ & $15.89\pm0.43$ & $2.31\pm0.05$ & $0.59\pm0.06$ & $16.87\pm0.43$ & 156/116 \\
93105-02-57-00 & 2009-11-20 & $0.52\pm0.03$ & $0.33\pm0.04$ & $15.70\pm0.39$ & $2.89\pm0.06$ & $0.80\pm0.07$ & $16.33\pm0.43$ & 121/116 \\
93105-02-58-00 & 2009-11-28 & $1.15\pm0.08$ & $0.27\pm0.06$ & $14.61\pm0.54$ & $4.26\pm0.10$ & $1.06\pm0.17$ & $11.98\pm0.73$ & 141/116 \\
93105-02-39-00 & 2010-01-11 & $0.50\pm0.05$ & $0.20\pm0.08$ & $17.20\pm0.80$ & $2.48\pm0.09$ & $0.94\pm0.17$ & $14.97\pm1.01$ & 135/116 \\
93105-02-45-00 & 2010-01-16 & $0.35\pm0.02$ & $0.28\pm0.06$ & $17.62\pm0.52$ & $2.20\pm0.07$ & $0.92\pm0.12$ & $16.23\pm0.66$ & 122/116 \\
93105-02-49-00 & 2010-01-25 & $0.61\pm0.06$ & $0.23\pm0.07$ & $15.56\pm0.71$ & $2.97\pm0.09$ & $0.91\pm0.14$ & $16.48\pm0.83$ & 129/116 \\
93105-02-51-00 & 2010-01-29 & $1.34\pm0.07$ & $0.29\pm0.04$ & $15.77\pm0.42$ & $4.28\pm0.07$ & $1.55\pm0.21$ & $10.55\pm0.61$ & 154/116 \\
95105-01-01-00 & 2010-02-08 & $0.70\pm0.04$ & $0.28\pm0.05$ & $15.90\pm0.47$ & $3.14\pm0.06$ & $1.14\pm0.12$ & $14.86\pm0.56$ & 138/113 \\ 
95105-01-02-00 & 2010-02-13 & $0.62\pm0.05$ & $0.18\pm0.07$ & $16.98\pm0.56$ & $2.98\pm0.08$ & $0.98\pm0.12$ & $14.71\pm0.68$ & 145/116 \\
95105-01-03-00 & 2010-02-20 & $1.76\pm0.08$ & $0.53\pm0.06$ & $16.44\pm0.91$ & $5.71^{\mathrm{a}}$ & $0.57\pm0.37$ & $11.10\pm1.99$ & 97.9/94 \\
95105-01-04-00 & 2010-02-27 & $1.44\pm0.09$ & $0.41\pm0.06$ & $15.86\pm0.53$ & $4.57\pm0.10$ & $1.98\pm0.42$ & $10.22\pm0.84$ & 117/116 \\
95105-01-05-00 & 2010-03-07 & $2.99\pm0.24$ & $0.49\pm0.10$ & $17.56\pm0.72$ & -$^{\mathrm{b}}$ & -          &    -           & 118/119 \\  
95105-01-06-00 & 2010-03-12 & $1.97\pm0.10$ & $0.14\pm0.05$ & $21.93\pm0.39$ & -$^{\mathrm{b}}$ & -          &    -           & 154/119 \\        
95105-01-07-00 & 2010-03-20 & $0.59\pm0.03$ & $0.33\pm0.05$ & $15.87\pm0.44$ & $2.93\pm0.06$ & $0.94\pm0.09$ & $15.94\pm0.50$ & 124/116 \\
95105-01-08-00 & 2010-03-27 & $0.53\pm0.03$ & $0.22\pm0.05$ & $18.43\pm0.48$ & $2.72\pm0.07$ & $1.03\pm0.12$ & $14.65\pm0.63$ & 119/116 \\
95105-01-09-00 & 2010-04-03 & $0.59\pm0.03$ & $0.26\pm0.05$ & $16.66\pm0.45$ & $2.90\pm0.06$ & $1.09\pm0.10$ & $15.26\pm0.53$ & 132/116 \\
95105-01-10-00 & 2010-04-10 & $0.95\pm0.11$ & $0.20\pm0.08$ & $17.68\pm0.87$ & $3.36\pm0.11$ & $1.64\pm0.40$ & $12.10\pm1.24$ & 109/116 \\
95105-01-11-00 & 2010-04-22 & $1.61\pm0.11$ & $0.35\pm0.06$ & $16.60\pm0.54$ & $4.77\pm0.14$ & $2.15\pm0.61$ & $8.46\pm1.01$  & 162/116 \\
95105-01-12-00 & 2010-04-26 & $2.71\pm0.11$ & $0.67\pm0.07$ & $16.52\pm0.44$ & -$^{\mathrm{b}}$ & -          &    -           & 124/119 \\ 
95105-01-12-02$^{\mathrm{c}}$ & 2010-04-29 & $2.84\pm0.09$ & $0.77\pm0.08$ & $14.96\pm0.35$ & -$^{\mathrm{b}}$ & - & - & 118/116 \\
95105-01-12-01$^{\mathrm{d}}$ & &           &               &                &                  &            &                &         \\
95105-01-13-00$^{\mathrm{c}}$ & 2010-05-02 & $2.73\pm0.10$ & $0.73\pm0.09$ & $15.44\pm0.44$ & -$^{\mathrm{b}}$ &   -   & -    & 103/119 \\
95105-01-13-03$^{\mathrm{d}}$ & &           &               &                &                  &            &                &         \\
95105-01-13-02 & 2010-05-05 & $2.29\pm0.13$ & $0.58\pm0.08$ & $17.45\pm0.52$ & -$^{\mathrm{b}}$ & -          &    -           & 108/119 \\
95105-01-14-03 & 2010-05-07 & $2.70\pm0.18$ & $0.77\pm0.13$ & $15.82\pm0.69$ & -$^{\mathrm{b}}$ & -          &    -           & 111/119 \\
95105-01-14-00 & 2010-05-08 & $2.87\pm0.18$ & $0.73\pm0.13$ & $14.35\pm0.62$ & -$^{\mathrm{b}}$ & -          &    -           &  97/119 \\
95105-01-14-01 & 2010-05-10 & $2.25\pm0.11$ & $0.38\pm0.05$ & $19.07\pm0.40$ & -$^{\mathrm{b}}$ & -          &    -           & 111/119 \\
95105-01-14-02 & 2010-05-12 & $2.25\pm0.17$ & $0.35\pm0.07$ & $20.57\pm0.69$ & -$^{\mathrm{b}}$ & -          &    -           & 100/119 \\
95105-01-15-00 & 2010-05-14 & $1.62\pm0.08$ & $0.63\pm0.08$ & $16.35\pm0.53$ & $5.21\pm0.17$ & $2.48\pm0.74$ & $9.56\pm0.98$  & 107/116 \\
95105-01-15-01 & 2010-05-16 & $2.34\pm0.14$ & $0.40\pm0.06$ & $18.33\pm0.47$ & -$^{\mathrm{b}}$ & -          &    -           & 122/119 \\ 
95105-01-16-00 & 2010-05-22 & $3.18\pm0.18$ & $1.05\pm0.23$ & $13.25\pm0.76$ & -$^{\mathrm{b}}$ & -          &    -           & 150/119 \\
95105-01-16-01 & 2010-05-24 & $2.14\pm0.21$ & $0.69\pm0.17$ & $16.00\pm0.95$ & -$^{\mathrm{b}}$ & -          &    -           & 100/119 \\
95105-01-17-00 & 2010-05-29 & $0.95\pm0.13$ & $0.23\pm0.09$ & $17.10\pm1.01$ & $3.61\pm0.12$ & $1.67\pm0.48$ & $13.06\pm1.36$ & 126/116 \\
95105-01-17-01 & 2010-06-01 & $0.73\pm0.10$ & $0.19\pm0.10$ & $16.31\pm0.97$ & $3.18\pm0.14$ & $1.33\pm0.28$ & $14.83\pm1.19$ & 144/116 \\ 
95105-01-18-00 & 2010-06-04 & $0.50\pm0.03$ & $0.24\pm0.06$ & $18.15\pm0.53$ & $2.72\pm0.08$ & $0.84\pm0.10$ & $16.24\pm0.65$ & 115/116 \\
95105-01-18-01 & 2010-06-07 & $0.79\pm0.06$ & $0.43\pm0.08$ & $16.07\pm0.70$ & $3.67\pm0.18$ & $1.04\pm0.24$ & $13.57\pm1.01$ & 109/116 \\
95105-01-19-00$^{\mathrm{c}}$ & 2010-06-11 & $0.49\pm0.09$ & $0.11\pm0.13$ & $18.43\pm1.23$ & $2.60\pm0.14$ & $0.82\pm0.22$ & $15.03\pm1.58$ & 106/116 \\
95105-01-19-01$^{\mathrm{d}}$ & &           &               &                &                  &            &                &         \\
95105-01-19-02$^{\mathrm{c}}$ & 2010-06-12 & $0.44\pm0.05$ & $0.26\pm0.10$ & $18.78\pm0.99$ & $2.56\pm0.14$ & $0.75\pm0.22$ & $16.21\pm1.38$ & 100/116 \\
95105-01-19-03$^{\mathrm{d}}$ & &           &               &                &                  &            &                &         \\
\hline 
\end{tabular}
\begin{list}{}{}
\item[$^{\mathrm{a}}$] We fixed the frequency $\nu_h$ in order to fit $L_h$.
\item[$^{\mathrm{b}}$] $L_h$ is not needed in this fit.
\item[$^{\mathrm{c}}$] This observation and the following one were performed within 24 hours, hence we averaged and fitted power
spectra together.
\item[$^{\mathrm{d}}$] This observation and the previous one were performed within 24 hours, hence we averaged and fitted power
spectra together.
\end{list}
\end{table*}
%
\begin{table*}
\centering
\caption{- continued from Table \ref{tab:log_PDS_fit}.}
\label{tab:log_PDS_fit_bis}
\begin{tabular}{c c c c c c c c c}
\hline
           &                & \multicolumn{3}{c}{{\bf $L_b$}} & \multicolumn{3}{c}{{\bf $L_h$}} &                            \\
Obs. ID    & Date           & $\nu_b$ (Hz)  & $Q$ & rms ($\%$)     & $\nu_h$ (Hz)   & $Q$ & rms ($\%$)     & $\chi^2 /$d.o.f. \\
\hline
95105-01-19-04$^{\mathrm{e}}$ & 2010-06-15 & $0.34\pm0.03$ & $0.35\pm0.10$ & $17.91\pm0.99$ & $2.14\pm0.15$ & $0.60\pm0.17$ & $17.08\pm1.26$ & 79/70   \\
95105-01-20-00 & 2010-06-19 & $0.30\pm0.02$ & $0.33\pm0.06$ & $18.51\pm0.51$ & $1.93\pm0.07$ & $0.70\pm0.09$ & $16.20\pm0.62$ & 164/116 \\
95105-01-20-01 & 2010-06-24 & $0.36\pm0.03$ & $0.25\pm0.08$ & $17.97\pm0.71$ & $2.38\pm0.11$ & $0.73\pm0.18$ & $15.40\pm1.00$ & 97/116  \\
95105-01-21-00 & 2010-06-28 & $0.32\pm0.03$ & $0.14\pm0.09$ & $20.26\pm0.75$ & $2.11\pm0.08$ & $0.91\pm0.17$ & $16.06\pm1.01$ & 117/116 \\
95105-01-21-01 & 2010-07-01 & $0.32\pm0.02$ & $0.31\pm0.07$ & $19.91\pm0.59$ & $2.00\pm0.07$ & $0.96\pm0.16$ & $15.99\pm0.80$ & 114/116 \\
95105-01-22-00 & 2010-07-04 & $0.22\pm0.02$ & $0.26\pm0.08$ & $16.39\pm0.96$ & $1.32\pm0.08$ & $0.42\pm0.15$ & $16.06\pm1.17$ & 121/116 \\
95105-01-22-01 & 2010-07-08 & $0.28\pm0.05$ &      0        & $16.96\pm1.48$ & $1.62\pm0.29$ & $0.32\pm0.22$ & $13.92\pm2.04$ & 135/117 \\  
95105-01-23-00 & 2010-07-11 & $0.20\pm0.03$ & $0.24\pm0.13$ & $16.18\pm1.19$ & $1.35\pm0.11$ & $0.38\pm0.16$ & $16.53\pm1.34$ & 109/116 \\
95105-01-23-01 & 2010-07-14 & $0.15\pm0.02$ & $0.37\pm0.17$ & $13.09\pm1.18$ & $1.06\pm0.16$ &       0       & $17.18\pm0.83$ & 131/117 \\
95105-01-24-00 & 2010-07-20 & $0.13\pm0.02$ & $0.40\pm0.20$ & $14.74\pm1.33$ & $1.08\pm0.29$ &       0       & $16.50\pm1.02$ & 131/117 \\
95105-01-25-00 & 2010-07-26 & $0.32\pm0.05$ &       0       & $20.70\pm0.71$ & -$^{\mathrm{b}}$ & -          &    -           & 118/120 \\ 
95105-01-25-01 & 2010-07-28 & $0.18\pm0.02$ &       0       & $21.01\pm0.63$ & -$^{\mathrm{b}}$ & -          &    -           & 111/120 \\

\hline 
\end{tabular}
\begin{list}{}{}
\item[$^{\mathrm{b}}$] $L_h$ is not needed in this fit.
\item[$^{\mathrm{e}}$] This observations has a spike in one bin of the light curve of PCU0. In order to fit the power spectrum,
we removed the bin.
\end{list}
\end{table*}
%
\begin{table*}
\centering
\caption{Best-fitting parameters of 59 energy spectra of Swift J1753.5-0127 extracted from RXTE observations performed during the
``failed transition''. The details on the extraction of the spectra and on the fitting procedures
are described in \S \ref{par:spectral_analysis}. For all the observations performed before December 2009
we made use of PCA and HEXTE data while for the remaining observations only PCA data were available.
Here we report the values of the parameters of the broken power law used to fit the spectra, as well as the unabsorbed fluxes in the 2-20 keV
energy range.
This Table continues in Table \ref{tab:log_fit_RXTE_spectra_2}.}
\label{tab:log_fit_RXTE_spectra_1}
\begin{tabular}{c c c c c c}
\hline
Obs. ID        & $\chi^2 /$d.o.f. & $\Gamma_1$    & $E_{break}$             & $\Gamma_2$       & Unabsorbed flux (erg/cm$^2$/s; 2-20 keV)   \\
\hline
\multicolumn{6}{c}{{\it PCA \& HEXTE}} \\
93105-02-36-00 & 122/104          & $1.65\pm0.01$ & $11.02^{+0.83}_{-0.63}$ & $1.48\pm0.02$    &   $1.38 \times 10^{-9}$   \\
93105-02-37-00 & 118/104          & $1.67\pm0.01$ & $9.92^{+0.90}_{-0.67}$  & $1.52\pm0.02$    &   $1.40 \times 10^{-9}$   \\          
93105-02-38-00 & 87/104           & $1.68\pm0.01$ & $10.39^{+0.68}_{-0.74}$ & $1.53\pm0.02$    &   $1.50 \times 10^{-9}$   \\
93105-02-40-00 & 130/104          & $1.71\pm0.01$ & $10.41^{+0.80}_{-0.67}$ & $1.54\pm0.03$    &   $1.70 \times 10^{-9}$   \\
93105-02-41-00 & 97/104           & $1.69\pm0.01$ & $9.93^{+0.62}_{-0.55}$  & $1.54^{+0.01}_{-0.02}$ &   $1.64 \times 10^{-9}$ \\
93105-02-42-00 & 86/104           & $1.68\pm0.01$ & $10.05^{+0.68}_{-0.67}$ & $1.53\pm0.02$    &   $1.47 \times 10^{-9}$       \\
93105-02-43-00 & 114/104          & $1.68\pm0.01$ & $10.95^{+0.67}_{-0.70}$ & $1.53\pm0.02$    &   $1.58 \times 10^{-9}$       \\
93105-02-44-00 & 107/104          & $1.69\pm0.01$ & $10.34^{+0.72}_{-0.91}$ & $1.56\pm0.02$    &   $1.69 \times 10^{-9}$       \\
93105-02-46-00 & 94/104           & $1.72\pm0.02$ & $10.14^{+1.42}_{-0.77}$ & $1.56\pm0.02$    &   $1.73 \times 10^{-9}$       \\
93105-02-47-00 & 122/104          & $1.74\pm0.01$ & $9.69\pm0.66$           & $1.62\pm0.01$    &   $1.93 \times 10^{-9}$       \\
93105-02-48-00 & 116/104          & $1.77\pm0.01$ & $9.62^{+0.47}_{-0.45}$  & $1.61\pm0.01$    &   $2.00 \times 10^{-9}$       \\
93105-02-50-00 & 130/104          & $1.81\pm0.01$ & $10.00^{+0.68}_{-0.53}$ & $1.67\pm0.01$    &   $2.21 \times 10^{-9}$       \\
93105-02-52-00 & 132/104          & $1.74\pm0.01$ & $10.64^{+0.92}_{-0.93}$ & $1.59\pm0.02$    &   $1.89 \times 10^{-9}$       \\
93105-02-54-00 & 80/104           & $1.82\pm0.01$ & $9.78^{+0.61}_{-0.58}$  & $1.69\pm0.01$    &   $2.12 \times 10^{-9}$       \\
93105-02-55-00 & 129/104          & $1.87\pm0.01$ & $9.84^{+0.55}_{-0.53}$  & $1.71\pm0.01$    &   $2.18 \times 10^{-9}$       \\
93105-02-56-00 & 141/104          & $1.90\pm0.01$ & $11.31^{+0.53}_{-0.48}$ & $1.68\pm0.02$    &   $2.16 \times 10^{-9}$       \\
93105-02-57-00 & 115/104          & $1.98\pm0.01$ & $10.60^{+0.40}_{-0.37}$ & $1.73\pm0.02$    &   $2.13 \times 10^{-9}$       \\
93105-02-58-00 & 128/104          & $2.17\pm0.01$ & $10.78^{+0.33}_{-0.31}$ & $1.84\pm0.02$    &   $1.97 \times 10^{-9}$       \\
\hline
\multicolumn{6}{c}{{\it PCA only}} \\
93105-02-39-00 & 48/60            & $1.93\pm0.01$ & $11.11^{+0.53}_{-0.46}$ & $1.66\pm0.02$    &   $1.84 \times 10^{-9}$       \\
93105-02-45-00 & 72/60            & $1.89\pm0.01$ & $9.85^{+0.62}_{-0.44}$  & $1.68\pm0.02$    &   $1.79 \times 10^{-9}$       \\
93105-02-49-00 & 62/60            & $1.99\pm0.01$ & $10.62^{+0.60}_{-0.45}$ & $1.75\pm0.02$    &   $1.82 \times 10^{-9}$       \\
93105-02-51-00 & 78/60            & $2.14\pm0.01$ & $11.32^{+0.57}_{-0.51}$ & $1.82\pm0.03$    &   $1.65 \times 10^{-9}$       \\
95105-01-01-00 & 61/60            & $2.01\pm0.01$ & $10.84^{+0.56}_{-0.49}$ & $1.80\pm0.02$    &   $1.75 \times 10^{-9}$       \\
95105-01-02-00 & 61/60            & $2.00\pm0.01$ & $10.21^{+0.44}_{-0.38}$ & $1.75\pm0.02$    &   $1.74 \times 10^{-9}$       \\
95105-01-03-00 & 89/60            & $2.21\pm0.01$ & $12.64^{+0.84}_{-0.77}$ & $1.69^{+0.06}_{-0.08}$ & $1.34 \times 10^{-9}$ \\
95105-01-04-00 & 73/60            & $2.20\pm0.01$ & $11.54^{+0.49}_{-0.41}$ & $1.78^{+0.03}_{-0.04}$ & $1.53 \times 10^{-9}$ \\
95105-01-05-00$^{\mathrm{a}}$ & 72/58 & $2.25^{+0.01}_{-0.03}$ & $11.65^{+0.58}_{-0.48}$ & $1.66^{+0.05}_{-0.06}$ & $1.05 \times 10^{-9}$ \\
95105-01-06-00 & 65/60            & $2.22\pm0.01$ & $12.08^{+0.55}_{-0.50}$ & $1.69\pm0.05$    &   $1.23 \times 10^{-9}$       \\
95105-01-07-00 & 61/60            & $1.97\pm0.01$ & $10.75^{+0.61}_{-0.52}$ & $1.74\pm0.02$    &   $1.67 \times 10^{-9}$       \\
95105-01-08-00 & 43/60            & $1.93\pm0.02$ & $10.56^{+0.55}_{-0.56}$ & $1.74\pm0.02$    &   $1.56 \times 10^{-9}$       \\
95105-01-09-00 & 66/60            & $1.95\pm0.01$ & $13.45^{+1.69}_{-0.92}$ & $1.50^{+0.13}_{-0.06}$ &  $1.54 \times 10^{-9}$  \\
95105-01-10-00 & 64/60            & $2.02\pm0.01$ & $11.95^{+0.96}_{-0.77}$ & $1.74^{+0.03}_{-0.04}$ &   $1.52 \times 10^{-9}$ \\
95105-01-11-00 & 83/60            & $2.18\pm0.01$ & $11.40^{+0.69}_{-0.61}$ & $1.79^{+0.04}_{-0.05}$ &   $1.33 \times 10^{-9}$ \\
95105-01-13-02 & 52/60            & $2.24\pm0.02$ & $11.47^{+1.02}_{-0.91}$ & $1.77^{+0.08}_{-0.10}$ &  $9.37 \times 10^{-10}$ \\
95105-01-14-03 & 68/60            & $2.63^{+0.17}_{-0.08}$ & $6.45^{+0.58}_{-0.53}$  & $2.01\pm0.04$ &  $9.58 \times 10^{-10}$ \\
95105-01-14-02 & 49/60            & $2.21^{+0.02}_{-0.01}$ & $10.73^{+0.55}_{-0.48}$ & $1.84\pm0.04$ &   $1.15 \times 10^{-9}$ \\
95105-01-15-00 & 41/60            & $2.18\pm0.02$ & $11.39^{+0.78}_{-0.56}$ & $1.78^{+0.05}_{-0.06}$ &   $1.15 \times 10^{-9}$ \\
95105-01-15-01 & 45/60            & $2.19^{+0.02}_{-0.01}$ & $12.76^{+0.95}_{-0.93}$ & $1.69^{+0.08}_{-0.09}$ & $1.08 \times 10^{-9}$ \\
95105-01-16-01 & 56/60            & $2.31\pm0.02$ & $10.26^{+0.55}_{-0.50}$ & $1.82^{+0.05}_{-0.06}$ &  $9.00 \times 10^{-10}$ \\
95105-01-17-00 & 35/60            & $2.04\pm0.01$ & $11.24^{+0.59}_{-0.65}$ & $1.75\pm0.03$     &   $1.36 \times 10^{-9}$      \\
95105-01-17-01 & 55/60            & $2.01\pm0.01$ & $10.68^{+0.63}_{-0.57}$ & $1.74\pm0.03$     &   $1.38 \times 10^{-9}$      \\
95105-01-18-00 & 48/60            & $1.91\pm0.01$ & $13.87^{+0.85}_{-1.67}$ & $1.59^{+0.07}_{-0.05}$ &   $1.31 \times 10^{-9}$ \\
95105-01-18-01 & 43/60            & $2.02\pm0.01$ & $13.16^{+0.76}_{-0.83}$ & $1.61\pm0.06$     &   $1.30 \times 10^{-9}$      \\
95105-01-19-00 & 53/60            & $1.91\pm0.02$ & $11.57^{+0.95}_{-0.75}$ & $1.62^{+0.04}_{-0.05}$ &   $1.31 \times 10^{-9}$ \\
95105-01-19-01 & 47/60            & $1.89\pm0.02$ & $13.17^{+2.03}_{-1.78}$ & $1.63^{+0.08}_{-0.09}$ &   $1.30 \times 10^{-9}$ \\
95105-01-19-02 & 52/60            & $1.88^{+0.02}_{-0.03}$ & $12.06^{+6.33}_{-2.13}$ & $1.70^{+0.06}_{-0.32}$ & $1.27 \times 10^{-9}$ \\
95105-01-19-03 & 44/60            & $1.92^{+0.03}_{-0.04}$ & $8.91^{+1.41}_{-0.78}$ & $1.74^{+0.03}_{-0.04}$  & $1.32 \times 10^{-9}$ \\
95105-01-19-04 & 68/60            & $1.85\pm0.02$ & $11.38^{+1.79}_{-0.93}$ & $1.60^{+0.04}_{-0.08}$ &   $1.26 \times 10^{-9}$ \\
95105-01-20-00 & 66/60            & $1.82^{+0.03}_{-0.01}$ & $15.20^{+1.61}_{-3.19}$ & $1.41\pm0.10$ &   $1.20 \times 10^{-9}$ \\
95105-01-20-01 & 38/60            & $1.87\pm0.01$ & $11.50^{+0.91}_{-0.60}$ & $1.54^{+0.04}_{-0.05}$ &   $1.28 \times 10^{-9}$ \\
\hline
\end{tabular}
\begin{list}{}{}
\item[$^{\mathrm{a}}$] A disc-blackbody component ({\it diskbb} with temperature $kT = 0.41^{+0.11}_{-0.06}$ keV)
needs to be added in order to get a statistically acceptable fit. However, the normalization of the {\it diskbb} is not measured accurately
(the ratio between the disc normalization and its negative error at $1 \sigma$ is smaller than 3)
\end{list}
\end{table*}
%
\begin{table*}
\centering
\caption{- continued from Table \ref{tab:log_fit_RXTE_spectra_1}}
\label{tab:log_fit_RXTE_spectra_2}
\begin{tabular}{c c c c c c}
\hline
Obs. ID        & $\chi^2 /$d.o.f. & $\Gamma_1$    & $E_{break}$             & $\Gamma_2$   &    Unabsorbed flux (erg/cm$^2$/s; 2-20 keV) \\
\hline
95105-01-21-00 & 39/60            & $1.84^{+0.02}_{-0.01}$ & $11.34^{+1.84}_{-1.12}$ & $1.65^{+0.03}_{-0.06}$ & $1.22 \times 10^{-9}$ \\
95105-01-21-01 & 37/60            & $1.84^{+0.02}_{-0.01}$ & $10.59^{+0.68}_{-0.61}$ & $1.61\pm0.03$ & $1.22 \times 10^{-9}$ \\
95105-01-22-00 & 54/60            & $1.78\pm0.02$ & $9.34^{+1.09}_{-0.71}$ & $1.63^{+0.02}_{-0.03}$  & $1.09 \times 10^{-9}$ \\
95105-01-22-01 & 57/60            & $1.76\pm0.03$ & $9.98^{+2.27}_{-1.09}$  & $1.55^{+0.03}_{-0.08}$ & $1.05 \times 10^{-9}$ \\
95105-01-23-00 & 40/61            & $1.74\pm0.01$ & $14.54^{+3.26}_{-3.06}$ & $1.44\pm0.04$          & $1.04 \times 10^{-9}$ \\ 
95105-01-23-01 & 53/61            & $1.74\pm0.02$ & $9.87^{+0.89}_{-0.90}$ & $1.58\pm0.02$           & $9.67 \times 10^{-10}$ \\
95105-01-24-00 & 41/60            & $1.71\pm0.01$ & $15.18^{+3.83}_{-1.86}$ & $1.42^{+0.22}_{-0.24}$ & $8.03 \times 10^{-10}$ \\
95105-01-25-00 & 52/60            & $1.75^{+0.03}_{-0.02}$ & $9.23^{+0.84}_{-0.73}$ & $1.51\pm0.03$  & $8.51 \times 10^{-10}$ \\
95105-01-25-01 & 49/60            & $1.73^{+0.02}_{-0.03}$ & $10.60^{+1.29}_{-0.72}$ & $1.43^{+0.04}_{-0.06}$ & $8.42 \times 10^{-10}$ \\
\hline
\end{tabular}
\end{table*}
%
\begin{table*}
\centering
\caption{Best-fitting parameters of 8 energy spectra from simultaneous RXTE/PCA plus Swift/XRT observations of Swift J1753.5-0127 performed during the
``failed transition''. The details on the extraction of the spectra and on the fitting procedures
are described in \S \ref{par:spectral_analysis}. Here we report the values of the parameters of the
absorption $N_H$, the thermal component {\it diskbb} and the broken power law used to fit the spectra. 
In the last column we show the unabsorbed fluxes for the energy band 2-20 keV.
In order to fit all the information in one row, we truncated the RXTE obs. ID. The correct obs. ID can be obtained by adding ``95105-'' at the beginning (e.g. 95105-01-12-00).}
\label{tab:log_fit_Swift_spectra}
\begin{tabular}{c c c c c c c c c c}
\hline
\multicolumn{2}{c}{Obs. ID}  & Component:       & {\it wabs} & \multicolumn{2}{c}{{\it diskbb}} &  \multicolumn{3}{c}{{\it bknpower}}  &   \\
RXTE &  Swift    & $\chi^2 /$d.o.f. & $N_H^{\mathrm{a}}$ & kT (keV) & $R_{in}^{\mathrm{b}}\,\sqrt{cos \, \theta}$ & $\Gamma_1$ & $E_{break}$ & $\Gamma_2$ & Flux$^{\mathrm{c}}$ \\ 
\hline
01-12-00 & 00031232008 & 357/426 & $0.283^{+0.014}_{-0.004}$ & $0.273^{+0.006}_{-0.007}$ & $199^{+44}_{-45}$$^{\mathrm{d}}$ & $2.24\pm0.02$ & $11.95^{+0.74}_{-0.88}$ & $1.67^{+0.09}_{-0.10}$ & $9.02$ \\
01-12-02 & 00031232009 & 389/426 & $0.178^{+0.007}_{-0.010}$ & $0.415^{+0.004}_{-0.011}$ & $66^{+5}_{-3}$ &$2.28^{+0.03}_{-0.02}$&$10.97^{+0.62}_{-0.55}$ & $1.70^{+0.07}_{-0.08}$ & $8.50$ \\
01-12-01 & 00031232009 & 386/426 & $0.18^{+0.02}_{-0.01}$ & $0.41^{+0.01}_{-0.04}$ & $67^{+24}_{-4}$  & $2.28^{+0.02}_{-0.03}$ & $10.71^{+0.60}_{-0.52}$  & $1.73^{+0.09}_{-0.04}$ & $8.59$ \\
01-13-00 & 00031232010 & 381/426 & $0.272^{+0.010}_{-0.009}$ & $0.32^{+0.03}_{-0.02}$ & $132^{+20}_{-23}$$^{\mathrm{d}}$ & $2.32^{+0.01}_{-0.02}$ & $10.10^{+0.23}_{-0.38}$ & $1.79\pm0.02$  & $9.70$    \\
01-13-03 & 00031232011 & 397/426 & $0.209^{+0.006}_{-0.016}$ & $0.36\pm0.02$       & $94^{+13}_{-9}$  & $2.27^{+0.05}_{-0.02}$ & $10.81^{+0.46}_{-0.73}$ & $1.68^{+0.09}_{-0.06}$ & $10.5$ \\
01-14-00 & 00031232013 & 398/426 & $0.20\pm0.02$ & $0.34\pm0.02$          & $106^{+23}_{-13}$  & $2.30^{+0.04}_{-0.03}$ & $10.51^{+0.69}_{-0.66}$ & $1.74^{+0.07}_{-0.08}$ & $9.83$ \\
01-14-01 & 00031232014 & 326/426 & $0.21^{+0.03}_{-0.01}$ & $0.32^{+0.03}_{-0.04}$ & $115^{+61}_{-20}$$^{\mathrm{d}}$ & $2.16\pm0.02$   & $11.86^{+0.95}_{-1.01}$ & $1.76^{+0.07}_{-0.04}$ & $11.4$ \\
01-16-00 & 00031232018 & 369/426 & $0.23\pm0.01$ & $0.303^{+0.013}_{-0.007}$ & $144^{+20}_{-16}$ & $2.29\pm0.01$ & $13.18^{+0.91}_{-0.97}$  & $1.53^{+0.15}_{-0.17}$ & $8.82$ \\
\hline
\end{tabular}
\begin{list}{}{}
\item[$^{\mathrm{a}}$] The values of $N_H$ are in units of $\times 10^{22}$cm$^{-2}$.
\item[$^{\mathrm{b}}$] The inner disc radius (in kilometres) is related to the normalization $K$ of the {\it diskbb} component in this way:
$R_{in} = \frac{D_{10}}{\sqrt{cos \, \theta}} \times \sqrt{K}$, where $\theta$ is the disc inclination on the line of sight and $D_{10}$ is
the distance to the source (in units of 10 kpc). The distance to the source is $D > 7.2$ kpc (Zurita et al. 2008), hence we use $D = 8$ kpc.
\item[$^{\mathrm{c}}$] The value of the flux is in units of $\times 10^{-10}$ erg/cm$^2$/s. 
\item[$^{\mathrm{d}}$] For this fit the normalization of the {\it diskbb} is not measured accurately (the ratio between the disc normalization and its negative error
at $1 \sigma$ is smaller than 3).
\end{list}
\end{table*}
%
\label{lastpage}
\end{document}